\documentclass[]{article}
\usepackage[]{geometry}
\usepackage{graphicx}
\usepackage{epsfig}
\usepackage{amssymb,amsfonts,amsmath,amsthm}
\usepackage{mathrsfs}
\usepackage{epstopdf}
\usepackage{hyperref}
\usepackage{color}
\usepackage{natbib}
\usepackage{setspace}
\usepackage{enumitem}

\newtheorem{defi}{Definition}
\newtheorem{thm}{Theorem}

\newtheorem{axiom}{Axiom}

\newtheorem{prob}{Problem}
\newtheorem{sol}{Solution}

\pdfoutput=1

\begin{document}

\title{Chapter 3: The emergence of spacetime from causal sets}
\author{Christian W\"uthrich and Nick Huggett\thanks{This is a chapter of the planned monograph \emph{Out of Nowhere: The Emergence of Spacetime in Quantum Theories of Gravity}, co-authored by Nick Huggett and Christian W\"uthrich and under contract with Oxford University Press. More information at \url{www.beyondspacetime.net}. The primary author of this chapter is Christian W\"uthrich (christian.wuthrich@unige.ch). This work was supported financially by the ACLS and the John Templeton Foundation (the views expressed are those of the authors not necessarily those of the sponsors). We wish to thank 
Fay Dowker for correspondence.}}
\date{7 September 2020}
\maketitle

\label{ch:causets2}


\tableofcontents
\

\section{A statement of the problem}
\label{sec:prob}

The conclusion of the last chapter denied that a non-spatiotemporal world is impossible: according to causal set theory (CST), it is physically (and hence presumably metaphysically) possible that the world is not spatiotemporal. After having introduced CST in the previous chapter and illustrating how at least space and arguably spacetime disappears in CST, we set out in this chapter to show how relativistic spacetime may emerge from a causal set, or `causet'.

The immediate question before us concerns the relation between causets and spacetimes: how do relativistic spacetimes emerge from the fundamental structures, i.e., causets, in CST? Since CST, like any theory of quantum gravity, is supposed to be a theory of gravity more fundamental than GR, and since relative fundamentality in physics typically parallels scales of energy and size, we are looking to understand the low-energy, large-scale limit of CST. Thus, we are at the same time interested in the relationship between two theories---CST and GR in the present case---, as well as between the entities or structures postulated by these two theories---presently, causets and relativistic spacetimes. Whenever there is no danger of conflating the two---and this will normally be the case---, we will often not explicitly indicate which is the present topic. As we will mostly be concerned with the relationship between causets and relativistic spacetimes, we will be asking, more specifically, \emph{whether we can precisely state}, i.e., with full mathematical and philosophical rigour, \emph{the necessary (\S\ref{sec:necconds}) and sufficient (\S\ref{sec:sptfunctionalism}) conditions for generic causets to give rise to physically reasonable relativistic spacetimes}. Whatever the answer to this question, it better not merely deliver necessary and sufficient conditions for mathematical derivations, but supplement them with demonstrations of the `physical salience' of these derivations. The argument in \S\ref{sec:sptfunctionalism} for the sufficiency of the conditions listed in \S\ref{sec:necconds} relies on a defence of spacetime functionalism. 

An important insight of this chapter will be that assuring the physical salience of the relation between causets and spacetimes involves addressing foundational and even metaphysical issues. In particular, the philosophy of time will assume an important role in this undertaking. Consequently, this chapter contains sections on the implications of CST for the metaphysics of time (\S\ref{sec:becoming}) and a (novel) form of non-locality implied by the discreteness of the fundamental structure together with the demand for (at least emergent) Lorentz symmetry (\S\ref{sec:lorentz}). With this in place, let us start.

\section{Identifying necessary conditions}
\label{sec:necconds}

\subsection{The basic setup}
\label{ssec:setup}

The restriction to \emph{physically reasonable} spacetimes is supposed to leave room for CST to correct GR regarding which spacetimes are physically possible: GR is a notoriously permissive theory, and we would generally expect the more fundamental theory to rule out some relativistic spacetimes as physically impossible. If this were to happen, the question of what it is for a relativistic spacetime to be physically reasonable would find some serious traction. Before some theory of quantum gravity is established though, answers to this question merely trade in human intuitions and the prejudices of physicists. Since these preconceptions are all trained in our manifestly spatiotemporal world, relying on them in the quest for a fundamental theory of gravity becomes deeply problematic.\footnote{For an argument of how physicists' intuitions about what it is for a spacetime to be physically reasonable can mislead us, e.g., in the case of singular or non-globally hyperbolic spacetimes, see \citet{smewut11}, \citet{timeandagain}, and \citet{Dobo2017}.} 

In this sense, we will use the qualification of `physical reasonableness' merely as a reminder that we ought not to expect a quantum theory of gravity to return all sectors of Einstein's field equation, i.e., all spacetimes which are admissible by GR's lights. Thus, CST may only reproduce some, but not all, sectors of GR. For instance, as we have seen in chapter 2
, the antisymmetry assumed in the partial ordering rules out causal loops at the fundamental level, and so may prohibit spacetimes with closed timelike curves.\footnote{But cf.\ \citet{wut20}.} Furthermore, models in GR with high energy or matter density may be eliminated by underlying `quantum effects', some sectors of GR may violate energy conditions in ways inconsistent with CST. 

Conversely, we seek to understand how spacetime arises from \emph{generic} causets. On the one hand, it would be unreasonable to expect that all causets give rise to a spacetime, or even just nearly so. There will be causets without anything like an appropriately spatiotemporal structure. That such `pathological' cases will also satisfy the demands of the theory on a causet, and so qualify as physically possible, is one of the main reasons to consider the theory, in general, as non-spatiotemporal. On the other hand, we would want `most' causets, or at least `many' of them to give rise to spacetimes, for otherwise we might ask ourselves why we were so incredibly lucky to be born in a spatiotemporal world when `most'  of them are not so.\footnote{Arguably, we \emph{couldn't} have been born in a world best described by a non-spatiotemporal causet, and so a straightforward anthropic argument should eliminate any astonishment as to why our world is spatiotemporal, given our existence. However, this leaves bewildering the issue of why the world came to be spatiotemporal in the first place.} Clearly, more will need to be said about this---and more will be said instantly---, but we hope that the intuition behind demanding genericity is reasonably clear for now.

\begin{figure}[ht]
\centering
\epsfig{figure=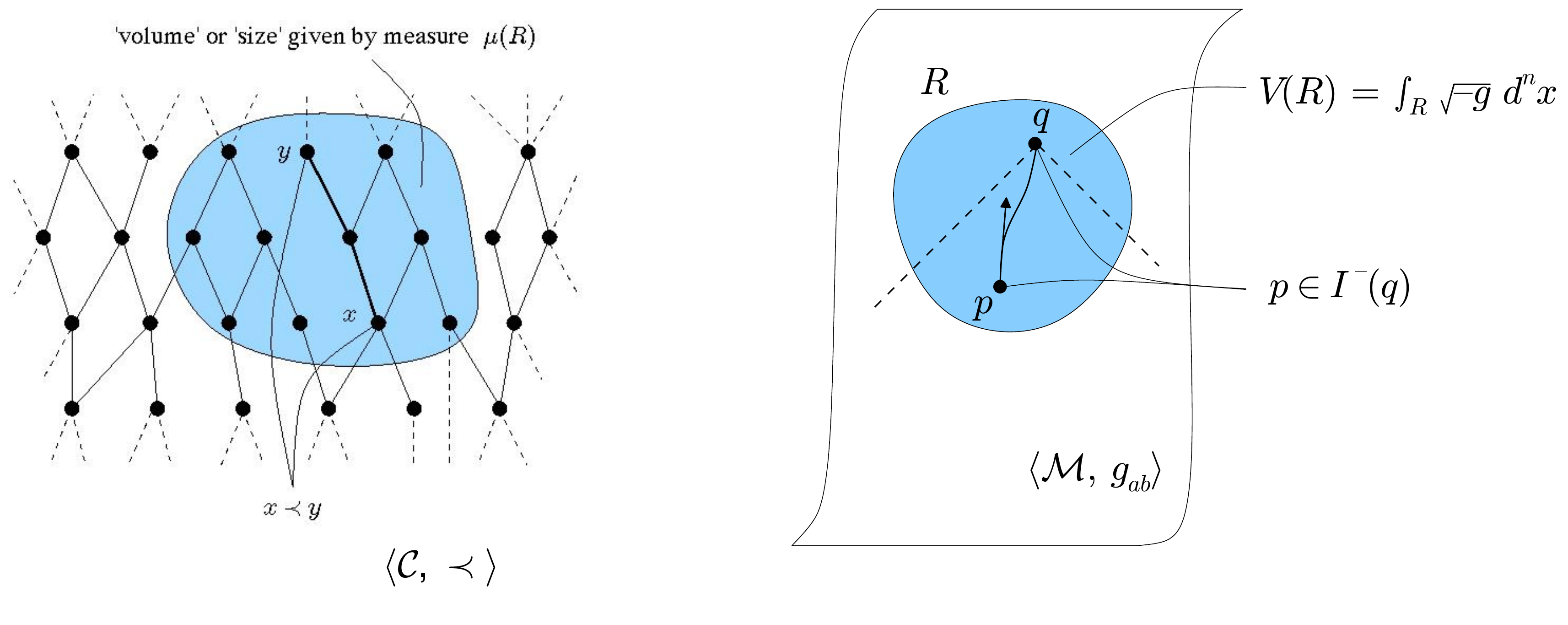,width=\linewidth}
\caption{\label{fig:emergingspt2} The relata.}
\end{figure}

Thus, we wish to relate generic causets to physically reasonable spacetimes. More specifically, we wish to see how the properties and structure of a part of a causet can give rise to the geometric and topological structure of a region of a relativistic spacetime. The basic idea is captured in figure \ref{fig:emergingspt2}. 
Importantly, it needs to be demonstrated how the relationship between features of causets and properties of spacetimes is not just mathematically definable, but physically salient. The method to accomplish this is not by staring at causets and ruminating over which of its parts are physically salient; instead one identifies the physically relevant properties of relativistic spacetimes and determines how physical salience percolates down to the fundamental structure: one tracks physical salience from the top down. But let us start with the more formal aspects of the relationship, hoping that the rest will then become more intelligible. 

Under what circumstances do causets give rise to relativistic spacetimes? A first necessary condition is that the causets are `sufficiently large' (``at least $10^{130}$'' elements, according to \citet{bomeal87}). The reason for this demand is evident: if it is too small, the notion of what it is approximated by 
at large scales does not make sense. Thus, let us eliminate all causets, which are not sufficiently large. Unfortunately, there is no hope that `sufficiently large' causets will generically give rise to spacetimes, due to the following problem:\footnote{This problem is what \citet[211]{smo06} calls the ``inverse problem'' and is often referred to as the ``entropy problem'' in the literature (see, for example, \citet{Brightwell2009}, \citet[188]{Dribus2017}, \citet[50]{Surya2019}), in reference to the fact from statistical physics that the large-scale behaviour of a physical system may be characterized in terms of the multiplicities of microscopic states sharing their macroscopic behaviour.}
\begin{prob}\label{prob:kr}
Almost all `sufficiently large' causets permitted by the kinematic axioms of chapter 2 
are so-called `Kleitman-Rothschild orders', i.e., discrete partial orders of only three `layers' or `generations' of elements. 
\end{prob}
`Kleitman-Rothschild orders' or \textit{KR orders} are partially ordered sets consisting in just three layers of elements such that no chain as defined in definition 8 in chapter 2 
consists in more than three elements, such that roughly half of the elements are in the middle layer and a quarter each in the top and bottom layers. Furthermore, each element of the middle layer is related by the order relation, on average, to half the elements of the top layer and to half the elements in the bottom layer.\footnote{For details, see \citet{klerot75}.} More precisely, \citet{klerot75} prove the following theorem:
\begin{thm}[Kleitman and Rothschild 1975]
Let $P_n$ denote the number of partial orders on a set of $n$ elements. Let $Q_n$ denote a special case in which $n$-element partially ordered sets are KR orders. Then 
\[
P_n = \left(1+\mathcal{O}\left(\frac{1}{n}\right)\right) Q_n.
\]
\end{thm}
In other words, in the limit of $n\rightarrow \infty$ of $n$-element partially ordered sets, almost all of them are KR orders. Why is this a problem? Because KR orders, if considered cosmological causets, would represent highly non-locally connected `universes' which `last' for a lousy three Planck times, i.e., roughly $10^{43}$s, during which they first double in size and then shrink by the same factor.\footnote{See figure 9 in \citet{Surya2019}.} If causets generically have the form of KR orders, then CST cannot offer a satisfactory answer to how they could give rise to anything like our world. It seems obvious that the vast majority of even sufficiently large causets cannot give rise to anything like spacetime as we know it. But how to tame the KR flood? Clearly, if the theory is conceived of as only comprising the kinematics described in the last chapter, it is too weak to clear out the KR weeds. Thus, what is needed are additional, and sufficiently restrictive laws. 

\subsection{Classical sequential growth dynamics to the rescue}
\label{ssec:dynamics}

This is precisely the motivation behind introducing a `dynamics':
\begin{sol}
`Dynamical' rules should be appended to the kinematic axioms such that those causets that satisfy these additional axioms are not befallen by 
the KR catastrophe.
\end{sol}
Accordingly, we impose `dynamical' laws in order to avoid the KR catastrophe and to restrict the vast set of kinematically possible causets to the physically reasonable models of the theory. By far the most popular proposal for causet dynamics is a law of sequential growth as introduced by \citet{Rideout1999}. The central idea of a sequential growth dynamics that a causet `grows' by the sequential addition of newly `born' events one by one to the future of already existing events. Thus, what grows is the number of elements, and it is assumed that the `birthing' of new elements is a stochastic physical process in the following sense: the dynamics specifies transition probabilities for evolving from one causet in $\Omega (n)$ to another one in $\Omega (n+1)$, where $\Omega (n)$ is the set of $n$-element causets. Since the growth always happens to the future of `already' existing events, the resulting causets are all finite towards the past or `past-finite', in the sense that they possess at least one minimal element. Only causets which could have been grown by a process consistent with the dynamical laws of sequential growth are then considered physically admissible. 

What dynamical laws ought to be postulated? They should capture our best guesses as to the conditions necessary to `produce' spacetime. In particular, they should thus be `natural'---i.e., physically salient---principles, not just arbitrary mathematical rules. Thus, these natural conditions encode the physical requirements on an acceptable dynamics. \citet{Rideout1999} impose four such requirements, of which we give here their own gloss:\footnote{The technically more precise statements can be found in \citet[\S III]{Rideout1999}. To give just an exemple, internal temporality demands, as stated above, just that if $x\preceq y$, then label$(x) \leq$ label$(y)$.}
\begin{axiom}[Internal temporality]\label{ax:temp}
``[E]ach element is born either to the future of, or unrelated to, all existing elements; that is, no element can arise to the past of an existing elment.'' (5)
\end{axiom}
\begin{axiom}[Discrete general covariance]\label{ax:gencov}
``[T]he `external' time in which the causets grows... is not meant to carry any physical information. We interpret this in the present context as being the condition that the net probability of forming any particular $n$-element causet $C$ is independent of the order of birth we attribute to its elements.'' (5f)
\end{axiom}
\begin{axiom}[Bell causality]\label{ax:bell}
``[E]vents occurring in some part of a causet $C$ should be influenced only by the portion of $C$ lying in their past.'' (6)
\end{axiom}
\begin{axiom}[Markov sum rule]\label{ax:markov}
``[T]he sum of the full set of transition probabilities issuing from a given causet [is] unity.'' (7)
\end{axiom}
Axioms \ref{ax:temp} and \ref{ax:gencov} are intended to jointly underwrite the idea central to classical sequential growth that although there is a birth order in the way the dynamics is described, it and the `time' in which it plays out has no physical significance. One would expect axiom \ref{ax:bell} to be violated in a quantum theory, which should naturally be non-local in that it admits Bell correlations, i.e., correlations among spacelike related events. For a classical theory, it seems unproblematic, and indeed felicitous, to postulate axiom \ref{ax:bell}, despite its being earmarked for being dropped in a quantum theory. Finally, axiom \ref{ax:markov}, required for any Markov process, just innocently assumes that for any given finite causet, there is exactly one way of possibly several in which it will in fact grow. Following, \citet{Rideout1999}, let us call a dynamical rule which complies with these four axioms a \textit{classical sequential growth dynamics}.

The four axioms can be thought to encapsulate the basic conditions any specific dynamic law must obey. Obviously, many dynamical laws specifying particular transition probabilities are conceivable. Yet a remarkable theorem by \citet{Rideout1999} shows that if the classical dynamics conforms to the four axioms \ref{ax:temp}--\ref{ax:markov}, then the dynamics is sharply constrained. In particular, it must come from a class of dynamics of sequential growth known as `generalized percolation'. Since only the general properties of this class matter for our purposes, let us introduce the simplest model of classical sequential growth which satisfies the Rideout-Sorkin theorem---namely, `transitive percolation'---a dynamics familiar in random graph theory. 

A simple way to understand this dynamics is to imagine a sequential order of event births, labelling that order using positive integers $1, 2, 3,...$ such that they are consistent with the causal order, i.e., if $x\preceq y$, then label$(x) \leq$ label$(y)$. The reverse implication does not hold because the dynamics at some label time may birth a spacelike-related event, not one for which $x\preceq y$. This is essentially the requirement of `internal temporality' introduced below. We begin with the causet's `big bang', the singleton set. Now when the second event is birthed, there are two possibilities: either the second event (labelled `2') causally succeeds the first event (labelled `1'), or it does not, i.e., $1\preceq 2$ or $\neg(1\preceq 2)$. Transitive percolation assigns some probability $p$ to the two events being causally linked and $1-p$ to the two events not being causally linked. The same holds for the third event, which has probability $p$ of being causally linked to 1 and $1-p$ of not being causally linked to 1, and probability $p$ that it is linked to 2 and $1-p$ that it is not linked to 2. 

In general, an alternative way to conceive of the dynamics is that when a new causet with $n+1$ events comes into being, it chooses a previously existing causet of $n$ events to be its ancestor with a certain probability. Thus, one can think of transitive percolation as involving two steps at each evolutionary stage from an $n$-element causet $\mathcal{C}_n$ to an $(n+1)$-element causet $\mathcal{C}_{n+1}$. First, select the subset of elements in $\mathcal{C}_n$ which will have a direct causal link to the newborn element. Second, add all ancestors implied by transitivity. In this way, the dynamics enforces transitive closure, so if, e.g., $1\preceq 2$ and $2\preceq 3$, then $1\preceq 3$. 

This second step means that the possible subsets of elements which can be ancestors is a subset of the power set of all elements: not all combinations of ancestors are generally permissible. Given an ancestor set, how is the probability determined? For each element in $\mathcal{C}_n$, the probability of it being part of a given set of ancestors is $p$. This determines the probability of the transition from a given $n$-element causet to a given $(n+1)-$element causet. For instance, the transition from a 2-chain ($1\preceq 2$) to a 3-chain ($1\preceq 2\preceq 3$) is the sum of the probability of picking both elements 1 and 2 of the 2-chain as ancestors and the probability of picking only the causally later element 2 as ancestor, since in this case, the second step (to ensure transitive closure), adds the first element 1 again. Thus, the probability of this transition is $p^2 + p(1-p)$, which happens to add up to $p$. 

\begin{figure}[ht]
\centering
\epsfig{figure=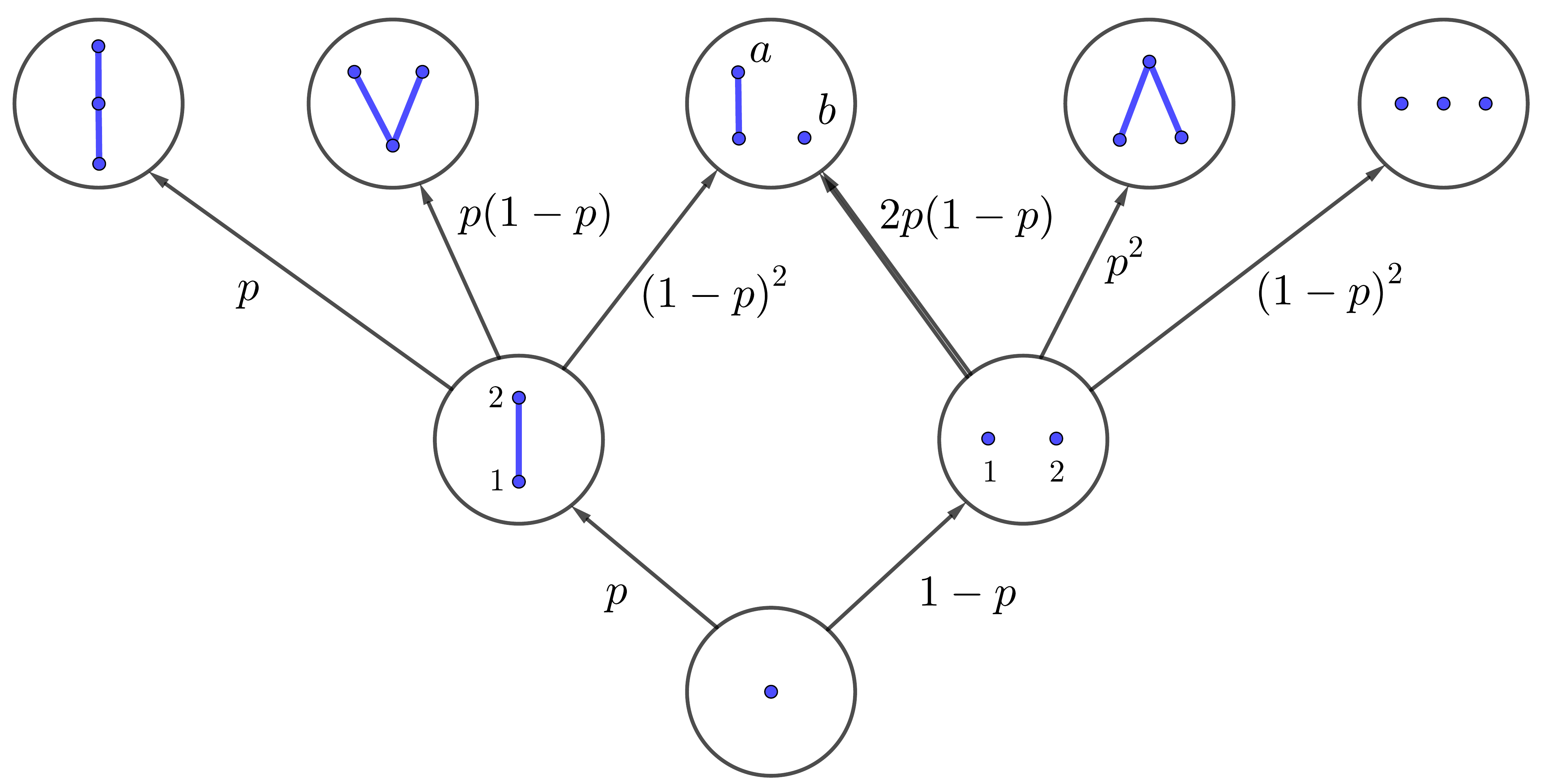,width=\linewidth}
\caption{\label{fig:transpercolation} The first three stages of transitive percolation dynamics.}
\end{figure}

A classical sequential growth dynamics can generally be though of as a `tree' of permissible transitions where each transition is assigned a probability consistent with the axioms above. The resulting tree is itself a partial order, where the ordering relation is not a causal relation intrinsic to a world, but instead a relation of permissible sequential growth. Figure \ref{fig:transpercolation} shows the first three stages with the transition probabilities exemplified by those of transitive percolation. 
In figure \ref{fig:transpercolation}, the thicker arrow indicates that there are two `ways' in which the transition from the 2-antichain to the 3-element causet with a chain of two elements and an isolated third element can occur: the third element can either be to the causal future of 1, or to the causal future of 2. Axiom \ref{ax:markov} requires that the transition probability must count both these possibilities; hence the factor of 2 in the transition probability. In turn, this naturally suggests that the dynamics presupposes a non-structuralist metaphysics of elementary events according to which these events have a primitive identity---a `haecceity', i.e., an identity which is independent of, and prior to, the causal relations they entertain to other events.\footnote{This raises the same metaphysical issue we have already encountered in the discussion of `distinguishing' causets in \S2.2,
 viz.\ whether elements are solely individuated by their structural position or whether they possess a haecceity beyond that. For further discussion of this point, see \citet{wut12}.}
 
That a careful consideration of the metaphysics of causets and their dynamics is needed can be directly seen when we seek reassurance that transitive percolation satisfies axioms \ref{ax:temp} through \ref{ax:markov}. While axioms \ref{ax:temp} and \ref{ax:bell} are directly and unproblematically built in, it may appear as if axiom \ref{ax:gencov} and axiom \ref{ax:markov} stand in tension. To repeat, the transition probability from the 2-antichain to the 3-element causet in the top middle of figure \ref{fig:transpercolation} must include the factor 2 for all three probabilities of the transitions emanating from the 2-antichain to sum to 1. However, axiom \ref{ax:gencov} demands that the product of the two transition probabilities from the ur-causet to the same 3-element causet along the two paths through the 2-chain and through the 2-antichain must be the same. Assuming $p\neq 0$, this cannot be the case if we maintain the factor 2. But the contradiction is only apparent: in order to check axiom \ref{ax:markov}, we must add up the probabilities of all possible ways in which the 2-antichain could evolve (including the two ways in which it can evolve to the 3-element causet of interest); in order to verify axiom \ref{ax:gencov}, however, we only need to multiply the probability to get to that 3-element causet from the same ur-causet along one path. Strictly speaking, there are three paths from the ur-causet to the 3-element causet at stake: one through the 2-chain, and two through the 2-antichain, and along all three paths, the probabilities must factor to the same product. And they do: they all factor to $p(1-p)^2$. 

There are further metaphysical assumptions built into the model of classical sequential growth dynamics. First, the probability of the `big bang' causet, i.e, the minimal element of the partial order depicted in figure \ref{fig:transpercolation}---the ur-causet---, is assumed to be 1. In other words, the dynamics presupposes that there is something rather than `nothing': the completely empty world containing no basal element at all is not a physical possibility. Second, as already mentioned, all causets which comply with this dynamics are past-finite and grow indefinitely into the future, at least as the transition probabilities are assumed to be non-zero. 

Third, as \citet{Rideout1999} declare right at the outset, and as has been reaffirmed by leading causet theorists since, it is their hope that this dynamical `birthing' of events, this sequentially growing `block universe', will turn out to underwrite and give rise to the ``phenomenological passage of time'', which ``is taken to be a manifestation of this continuing growth of the causet'' (2). This idea raises an immediate concern of conceptual inconsistency. On the one hand, physical time, and thus truly dynamical processes are supposed to only arise at the emergent level as an aspect of the emerging spacetime, and is at best only implicitly present in the causal structure of the causet. On the other hand, the rhetoric is infested by temporal locutions concerning the `growth' of causets, the `birthing' of events in `sequential' order following a `dynamical' law, suggesting the presence of time in a way that is metaphysically prior to the causet. Clearly, this tension has to be resolved, and the status of time will have to be settled. We will return to this and related metaphysical questions in section \ref{sec:becoming}. 

Let us add a final, but crucial point concerning sequential growth dynamics as a physical dynamical law for CST. Although it involves stochastic transitions between stages, there is nothing quantum in nature about the sequential growth dynamics---hence, it is known as \emph{classical} sequential growth dynamics. Just as discreteness postulated in the kinematics of the theory does not make the theory a quantum theory, neither does the stochasticity introduced in the dynamics. Alas, this is largely the state of the art as the philosopher studying CST finds it. Although there are incipient efforts to turn the theory into a full quantum theory such as those based on quantum measure theory \citep{Sorkin1997} or a quantization of classical sequential growth dynamics \citep{Gudder2014}, their incompleteness and inconclusiveness does not really permit a discussion of anything but the classical theory. Consequently, our conclusions are conditional on the proviso that this is just the classical theory so far and so will have to remain tentative. 

\subsection{Manifoldlikeness}
\label{ssec:manifold}

Transitive percolation or something like it ought to be able to turn the KR tide, at least as long as the probability $p$ does not vanish. Is adding a viable and physically motivated dynamical law thus everything required to understand the emergence of spacetime from causets? A priori, no: resulting causet are generally still not `manifoldlike' and so not yet candidates for giving rise to spacetime. 
\begin{prob}\label{prob:mani}
The conditions stated so far may not guarantee that causets give rise to manifolds of reasonably low dimensionality (i.e., significantly lower than $10^{130}$) or that it has Lorentzian signature or a reasonable causal structure.
\end{prob}

The discrete structures of causets and continuum Lorentzian spacetimes are mathematically rather different structures, as already noted. But in order for a causet to give rise to something that is at some scales well described by a relativistic spacetime, these structures cannot be too different. Thus, the second necessary condition after the requirement that candidate causets be `sufficiently large' is that they are `manifoldlike':
\begin{sol}\label{sol:mani}
Impose conditions such that, generically, causets $\langle \mathcal{C}, \prec\rangle$ are `manifoldlike'. 
\end{sol}
But what is it for a causet to be `manifoldlike' in the required sense? It means that it stands in an appropriate relation to spacetimes: 
\begin{defi}
A causet $\langle \mathcal{C}, \prec\rangle$ is \textit{manifoldlike} just in case it is `well approximated' by a relativistic spacetime $\langle\mathcal{M}, g_{ab}\rangle$. 
\end{defi}
If a causet is manifoldlike, we also say that the associated spacetime $\langle\mathcal{M}, g\rangle$ {\em approximates} the causet. Clearly, this definition, and hence solution \ref{sol:mani}, can only get any traction once we spell out what is meant by `well approximated'. To provide a rigorous understanding of what it would be for a causet and a spacetime to stand in the required relation turns out to be challenging and is the subject of ongoing research. There are at least two ways of approaching the problem. First, one could understand both causets and (distinguishing) Lorentzian manifolds as `causal measure spaces' and use a so-called \textit{Gromov-Hausdorff function} $d_{GH}(\cdot, \cdot)$ (or the `Gromov-Wasserstein function' if we have a probability measure) to give a measure of the closeness or similarity between such spaces.\footnote{This line is pursued by \citet{Bombetal12}, although that paper has been withdrawn and never been followed up, as far as we can tell. Sam Fletcher has given talks about this in 2013 developing this ansatz, but, to our knowledge, has not published this material.} Such a distance function relating any two causets, any two Lorentzian spacetimes, or any causet and any Lorentzian spacetime in terms of their relevant similarity  could provide a rigorous way to spell out the idea of `approximation'. 

But worked out in much more detail is a second approach, using the notion of a `faithful embedding' of a causet into a spacetime. The task here is to find a map $\varphi: \mathcal{C} \rightarrow \mathcal{M}$ such that the image looks like a relativistic spacetime, as shown in figure \ref{fig:emergingspt3}. More precisely, a spacetime approximates a causet if there exists a faithful embedding of the causet into the spacetime in the following sense:
\begin{figure}
\centering
\epsfig{figure=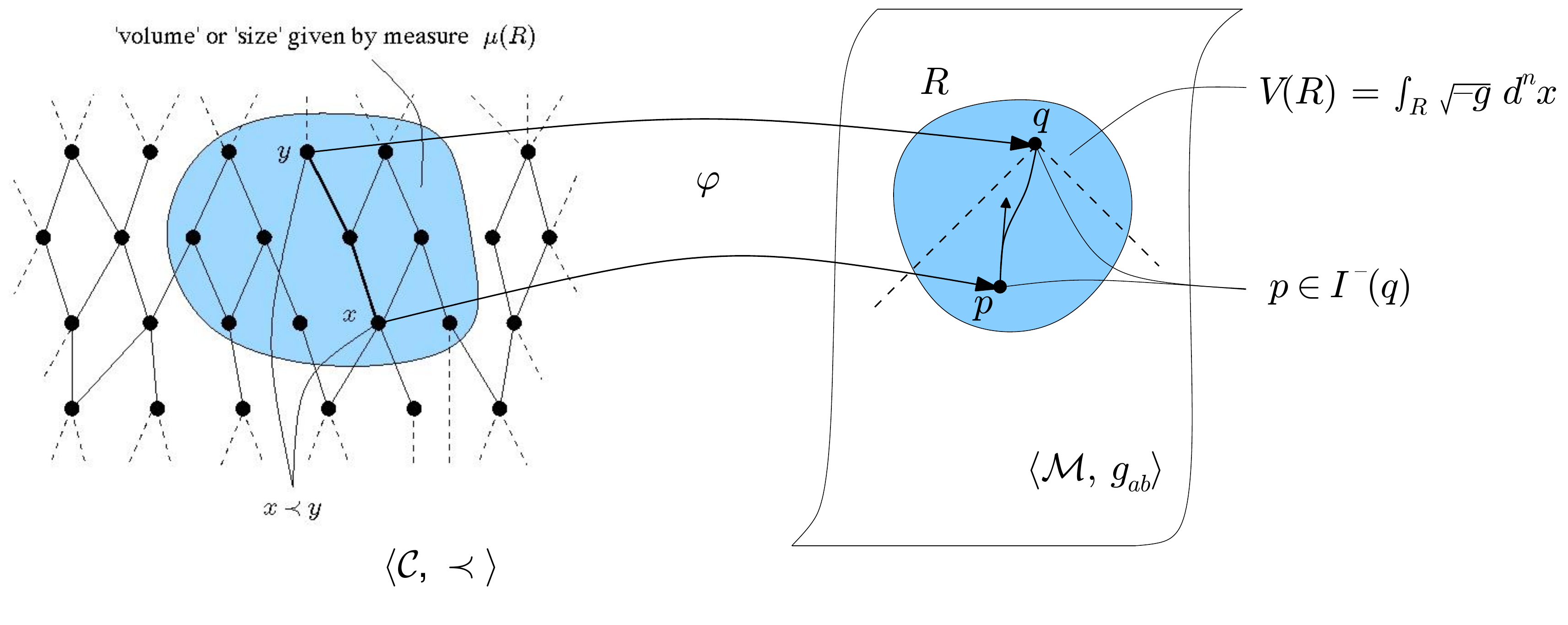,width=\linewidth}
\caption{\label{fig:emergingspt3} Faithful embedding (Figure from \citealt{lamwut18}).}
\end{figure}
\begin{defi}\label{def:faith}
A causet $\langle\mathcal{C}, \prec\rangle$ is said to {\em approximate} just in case there exists a `faithful embedding', i.e., a injective map $\varphi : \mathcal{C}\rightarrow\mathcal{M}$ such that
\begin{enumerate}
\item the causal relations are preserved, i.e.\ $\forall a, b\in\mathcal{C}, a \prec b$ iff $\phi(a) \in J^-(\phi(b))$;
\item $\varphi (\mathcal{C})$ is a `uniformly distributed' set of points in $\mathcal{M}$;
\item $\langle\mathcal{M}, g_{ab}\rangle$ does not have `structure' at scales below the mean point spacing.
\end{enumerate}
\end{defi}
Let us discuss the three conditions in turn.

The first condition requires that the causal structures of the causet and the spacetime to which it gives rise are isomorphic.\footnote{More precisely, the causet and the set of images of its events in the spacetime.} This requirement captures the idea that causal structure is fundamental and `percolates' up the scales in a precise sense. This demand seems acceptable for a fundamentally classical theory, such as the version of CST discussed here. Once this theory is replaced by a quantum CST, we cannot reasonably uphold this demand as we should expect quantum `fluctuations' of the causal structure rather than some determinate causal structure and so no untainted emergence of that structure. 

The second condition expresses the expectation that regions of causets of a similar number of elements give rise to spacetime regions of correspondingly similar volume. Thus, the mapping cannot be such that the image points of the causet elements are too dense and too sparse in the spacetime, resulting in a more or less uniform distribution. The intuition behind this condition is thus  clear, although its articulation obviously needs to be precisified.\footnote{One way in which it needs to be precisified is that the required uniformity must be with respect to the spacetime volume measure of $\langle \mathcal{M}, g\rangle$.}

The third condition, which will eventually also need a more rigorous formulation, insists on the absence of non-trivial structure in the spacetime between the uniformly distributed image points and so at scales so small that they could not possibly be captured by the causet. If this condition were violated, then the spacetime would have local features, such as a high or uneven curvature or a non-trivial topology in a very small region, for which the fundamental causet would simply lack the expressive power to include them. Suppose a given causet can be mapped into two two distinct spacetimes with the first two conditions being satisfied, such that one of the spacetimes contains non-trivial small-scale structure, while the other does not. Thus, the mapping into the second spacetime is faithful, but the one into the first is not. Since the point of the manifoldlikeness condition is to determine how we can conceive of the relation between causets and spacetime such that we can meaningfully state that the spacetime `emerges from' the causet, and since there is no reason to think that a causet can physically give rise to a spacetime with small-scale structure completely absent in the fundamental causet, there is no reason to validate the first mapping. 

A prototypical example to illustrate the relation between manifoldlike causets and relativistic spacetimes is the `Poisson sprinkling' of elements into two-dimensional (or higher-dimensional) Minkowski spacetime.\footnote{The image points of the causets in the spacetime cannot stand in a `regular', lattice-like structure since this would violate Lorentz invariance (see \S\ref{sec:lorentz}). The required `randomness' is naturally (though not uniquely) captured by a Poisson distribution.} Such a sprinkling selects events in Minkowski spacetime at the required density uniformly and at random and then imposes the unique partial ordering among them which is induced by the causal structure of Minkowski spacetime. Thus, Minkowski spacetime's causal structure is inherited by the causet and the first condition of a faithful embedding is satisfied by construction, as are the second and third.\footnote{For a useful illustration of Poisson sprinkling, see \citet[figure 1]{Dowker2013} and \citet[\S3, particularly figure 6]{Surya2019}.}

Notice that the way the relationship between fundamental causets and emergent spacetimes as assumed in this `sprinkling' is really the wrong way around: we started out with the spacetime given, and then sprinkled events into it and connected those which were causally related as given by the causal structure of the spacetime. Surely, this is putting the cart before the horse: the fundamental causet is prior to the relativistic spacetime, which is merely derivative. In this sense, what we ultimately want to understand is how a generic causet can give rise to the sorts of relativistic spacetimes we believe to be physically realistic. 

What remains unclear to date is what conditions need to be imposed in solution \ref{sol:mani} in order for the abiding causets to be manifoldlike. In CST, it is hoped that the dynamical conditions imposed to solve problem \ref{prob:kr} double up to also solve problem \ref{prob:mani}. In this sense, one would need to postulate one set of natural conditions or physically meaningful axioms, which solve both problems at once. Computer simulations fuel the hope that this may indeed be the case by providing evidence, e.g., that a significant class of dynamical laws closely related to transitive percolation yields a cyclically collapsing and re-expanding universe in which each era contains phases of exponential expansion approximated by de Sitter spacetime.\footnote{See \citet{Ahmed2010}. The class of dynamical laws is that of so-called `originary percolation', which is transitive percolation where the possibility of an element being born unrelated to any prior element is excluded.} 

\subsection{The Hauptvermutung}
\label{ssec:hauptvermutung}

The restriction to manifoldlike causets is surely a necessary condition for there to be any way in which a causet lends itself to the emergence of spacetime. But this in itself does not solve the challenge: GR is a notoriously permissive theory in that the number of spacetimes which satisfy the basic requirements of physical possibility imposed by GR, such as that the spacetime be Lorentzian and satisfies the Einstein field equation, is overwhelming. Many of these physically possible spacetimes are intuitively `unphysical' and so ruled out by physicists. This may be because they feature a pathological causal structure (containing, perhaps, closed timelike curves), an implausible topology, or a dubious curvature---or so we judge. In other words, we face the next problem:
\begin{prob}\label{prob:reason}
Given just how many `unphysical' spacetimes are physically possible according to GR, causets may generically give rise to `unphysical' relativistic spacetimes. 
\end{prob}
One might argue that this is not really a problem of CST, or indeed of any quantum theory of gravity, since it is already at the level of GR that we are faced with rampant unphysical solutions admitted by the theory. Although this is undisputedly the case, we could, and perhaps should, invert this reasoning: it is precisely \textit{because} GR is too permissive that we are looking to the more fundamental theory for guidance and hope that it will rule out many or most of these unphysical solutions. So this problem would be solved by the following solution:
\begin{sol}
Impose conditions such that qualifying causets are approximated by `physically reasonable' spacetimes.
\end{sol}
Evidently, this solution could only be implemented once we specify what a `physically reasonable' spacetime, a locution on which we certainly have an intuitive grasp, but which turns out to resist systematic and satisfactory treatment.\footnote{See \citet{Dobo2017} and \citet{manchak20} for further discussion of the notion of `physically reasonable' spacetimes.} Although a full explication is unlikely to come forward, we might hope to make some progress. For instance, one might think that the demand encoded in the kinematic axiom that the causal order be antisymmetric (cf.\ chapter 
2) ascertains that closed timelike curves cannot arise in the emerging spacetime. However, this implication only holds under some interpretations of the theory and must ultimately await its full articulation \citep{wut20}. Consequently, although the problem is real and the solution appears promising, its further elaboration is the task for another day.

Just as problem \ref{prob:mani}, it is hoped that no \textit{additional} conditions are needed and that the axioms postulated above would also suffice to generically deliver physically reasonable spacetime, whatever that is supposed to mean. But there is another reason why we would not want to impose any additional axioms just because they enforce our intuitions as to which spacetimes we take to be physically reasonable: part of the point of formulating a theory more fundamental than GR is for us to learn from it in which ways GR is false and needs to be corrected. In particular, this may apply to the range of physical possibilities which a fundamental theory more curtail much more restrictively than notoriously permissive GR. The point, however, would be to learn this from the fundamental theory, rather than not to conversely constrain that theory by the brute force of our antecedent intuitions. Thus, we take it that we have excellent reasons to reject problem \ref{prob:reason} as something we need to attend to at this point, all the while keeping in mind the issue as deserving revisiting once the theory is more fully understood.

In other words, the project is to determine, and to some extent to \textit{guess}, what the fundamental physics is which will produce structures which are generically approximated by relativistic spacetimes. 

But we are not quite done. Even if we thus arrive at the point at which judiciously chosen causets generically give rise to physically reasonable relativistic spacetimes, one further necessary condition must be imposed. A necessary condition for the spacetime to emerge is surely that it supervenes on the fundamental structure, to use the jargon of philosophers. In other words, there cannot be a difference at the level of the spacetime without there also being a difference at the level of the causet. The required covariance is asymmetric, as there can in general be differences at the level of causets that need not generate any difference for the emerging spacetime. We would want to think that the emergent spacetime asymmetrically depends on the causet. We can formulate this also as a problem:
\begin{prob}\label{pro:unique}
A given causet might be approximated by multiple relativistic spacetimes.
\end{prob}
This would indeed be a problem, since in this case, the fundamental structure would not uniquely determine the emergent large-scale structure. As \citet[313]{sor05} puts it: ``Implicit in the idea of a manifold approximating a causet is that the former is relatively unique; for if any two very different manifolds could approximate the same [causet] $C$, we'd have no objective way to understand why we observe one particular spacetime and not some very different one.'' The implicit condition then is something like this:
\begin{sol}\label{sol:haupt}
If a causet is approximated by a spacetime, then the approximating spacetime is `approximately unique'.
\end{sol}
Following Sorkin, this has become known as the `Hauptvermutung', or principal conjecture. Of course, this is but a template to fill in the specifics of the solution. In particular, what it means for the spacetime to be `approximately unique' will have to be specified. It should also be pointed out that it is hoped, or conjectured, that the satisfaction of the earlier conditions guarantees this solution such that no \emph{additional} condition ought to be imposed. In other words, the hope is that the Hauptvermutung can be established as a theorem of the theory as articulated so far and need not be added as an independent stipulation. In fact, to show that this is the case is the overarching goal of the research program (at least at the classical level). 

To date, the Hauptvermutung remains an unproven conjecture. In order to have any hope of proving it, a more precisely formulated proposition will be needed. Just how to accomplish that will depend on the chosen notion of `approximation'. Above, I identified two paths: one based on the Gromov-Hausdorff measure of similarity, and another in terms of a faithful embedding of causets into spacetimes. On the first research programme, the Hauptvermutung could be precisified as follows:
\begin{sol}[Hauptvermutung, Gromov-Hausdorff version]
If there exist two spacetimes $\langle \mathcal{M}, g_{ab}\rangle$ and $\langle \mathcal{M}', g'_{ab}\rangle$ such that $d_{GH}(\mathcal{C}, \mathcal{M}) < \epsilon$ and $d_{GH}(\mathcal{C}, \mathcal{M}') < \epsilon$ for some causet $\mathcal{C}$, then $d_{GH}(\mathcal{M}, \mathcal{M}') \\ < 2\epsilon$.
\end{sol}
Although the choice of some particular $\epsilon \in \mathbb{R}^+$ is rather arbitrary, something like the above condition could encapsulate the idea that a causet is approximated by a spacetime in an `approximately unique' sense. If, on the other hand, we followed the more standard path of articulating the relevant sense of approximation in terms of faithful embeddings, then the Hauptvermutung might be expressed differently:
\begin{sol}[Hauptvermutung, standard version]\label{sol:hauptvermutung}
If there exist two spacetimes $\langle \mathcal{M}, g_{ab}\rangle$ and $\langle \mathcal{M}', g'_{ab}\rangle$, which approximate a given causet $\mathcal{C}$ in that there exist faithful embeddings $\varphi: \mathcal{C} \rightarrow \mathcal{M}$ and $\varphi': \mathcal{C} \rightarrow \mathcal{M}'$, then they are `approximately isometric'.
\end{sol}
It might appear as if not much has been accomplished by replacing solution \ref{sol:haupt} by solution \ref{sol:hauptvermutung}, since we basically replaced the problematically vague notion of `approximate uniqueness' with an apparently equally problematically vague notion of `approximate isometry'. But we have made real progress: the first stab at solving problem \ref{pro:unique} just offered a general template of what, in rather general terms, would be needed to solve the problem, whereas the much more precise solution \ref{sol:hauptvermutung} reduces the problem to one of defining and defending a measure of approximate isometry between spacetimes. This is by no means a trivial problem, as is witnessed by the fact that it has so far resisted resolution. Such a resolution would presumably require that we identify the salient geometric properties of spacetime, introduce measures of similarity along the dimensions of the selected properties, and compound these measures into a score for approximate isometry. (Exact) isometry does not require any of the above; however, if the isometry is no longer `exact', and thus the metrics compared no longer identical in the relevant sense, the notion of (exact) isometry gets no traction, and the salient geometric features with respect to which the equivalence is supposed to be determined must be selected.\footnote{Two spacetimes $\langle \mathcal{M}, g_{ab}\rangle$ and $\langle \mathcal{M}', g'_{ab}\rangle$ are (exactly) \emph{isometric} just in case there exists a diffeomorphism $\phi:\mathcal{M} \rightarrow \mathcal{M}'$ such that $\phi_\ast (g_{ab}) = g'_{ab}$. This definition delivers the identity ``in the relevant sense'' invoked in the main text.} 


We shall leave it at this, as the point of this section is to sketch, only in general terms, what would be required in order to establish the emergence of spacetime in CST. Now suppose that the causet research programme offered satisfactory accounts of how to fill in the details in the above sketch and could thus be said to have fulfilled all necessary conditions listed. Of course, merely satisfying some necessary conditions gives us not guarantee that the work is jointly sufficient. And at the end of the day, we will want the assurance that the challenge of the emergence of spacetime is fully met, and this guarantee is only forthcoming if we convince ourselves that the necessary conditions we have discharged are also jointly sufficient.

\section{
The case for spacetime functionalism}
\label{sec:sptfunctionalism}

\subsection{The road to spacetime functionalism}
\label{ssec:road}

Before we enter the fray of the debate regarding the sufficiency of these conditions, let us pause to consider what would have been accomplished had we successfully discharged all necessary conditions. In fact, in this case, we could generically `derive'---i.e., systematically relate with full mathematical rigour---relativistic spacetimes from causets. No mean feat. But would this accomplishment conclude the project? Would we be done? Not according to the \emph{r\'esistance}, which believes that the conditions listed in the previous section are merely necessary but not jointly sufficient for the emergence of relativistic spacetimes from fundamental causets.

The \emph{r\'esistance} has different cells, which may be differently motivated. A first cell consists in \emph{primitive ontologists} who insist on fundamental local (and so a fortiori \emph{localizable}) beables in spacetime. According to this view, for a theory to qualify as a candidate physical theory describing a sector of our physical world and thereby accounting for some of our empirical data or, more broadly, for aspects of human experience, the theory must postulate an ontology of entities populating regions of spacetime. If the entities posited by a theory are not localizable, we seem to be losing a convenient way of parsing that which exists into a plurality of distinct entities: location offers a straightforward criterion for individuating objects. Although the loss of this criterion is already adumbrated by the non-locality of quantum physics, without spacetime at our side, how are we to dissect that which exists fundamentally into distinct and separate entities? Clearly, there are alternative ways of distinguishing entities, and so the criterion of spatiotemporal location commands no necessity. Admittedly though, for these alternative criteria to get traction at the fundamental level of a theory of quantum gravity may be tricky, and so a structuralist or monist stance may most naturally fit these cases.\footnote{Cf.\ \citet{wut12}, who argues for a structuralist interpretation of CST, and discusses the challenges of such an approach.} Even so, fundamental spatiotemporality is not necessary for parsing an ontology. Although the realist motivation of this camp is laudable, the particular form of realism demanded thus comes with an unduly narrow conception of fundamental ontology---one that quantum theories of gravity will be challenged to satisfy. 

Perhaps the concern is better expressed by a second cell, the \emph{anti-Pythagoreans} who fear a loss of the venerable distinction between physical stuff and abstracta, and object to the idea that the physical world is fundamentally mathematical or abstract by its nature. The distinction between concrete physical entities and abstract entities is thought to be threatened by the presumed non-spatiotemporality of a putative ontology of quantum gravity. On a standard demarcation, the concrete entities are taken to be those in spacetime, and the abstract ones those which are not. This criterion of demarcating the concrete from the abstract seems to suggest that the fundamental, non-spatiotemporal structures which make up our natural world must be abstract, perhaps mathematical. Alternatively, concrete physical entities have been characterized as those engaging in the causal commerce of the world. Worries of circularity concerning this criterion aside, attempts to explicate a notion of causal efficacy in the absence of spacetime may be thwarted by insurmountable difficulties \citep[\S4.2]{lamesf13}. Consequently, at least the most common ways of demarcating the physical from the mathematical are not apt in the present non-spatiotemporal context and it may be feared that the concerned theories are thus bound to vindicate the unwanted Pythagorean conclusion that our physical world is ultimately mathematical in nature. 

A third---related---cell is formed by those worried about the empirical coherence of theories without fundamental spacetime. All data ever collected for the empirical confirmation of physical theories seem, at their core, to be inextricably spatiotemporal: the coincidence of the tip of a pointer with a mark on a scale, the flashing of a particular light bulb, or the printout of a measurement outcome all involve physical objects or processes to be located in space and time. Thus, it seems as if the existence of spacetime is a precondition for empirical science. It appears as if theories of quantum gravity \emph{sans} spacetime deny the very conditions necessary for their empirical confirmation, and in this sense be `empirically incoherent' \citep{hugwut13}. Although empirical incoherence does not amount to any kind of impossibility, it would be most unfortunate for empirical science if we lived in a world in which the necessary conditions for such an enterprise could not be set in place. 

We believe that all these cells of the \emph{r\'esistance} can be put to rest if it could only be established that spacetime emerged in the appropriate limit or at the requisite scales. Showing how spacetime emerges would involve demonstrating how macroscopic (or indeed most microscopic) objects have location and are thus localizable. In this sense, to the extent to which the demand of primitive ontologists was reasonable, it would be met. Pythagoreanism would be averted, because although the fundamental ontology would neither be spatiotemporal nor directly causally efficacious, the fundamental structures would have been shown to directly connect to the spatiotemporal structures we associate with physical being. Presumably, the emergent spacetime would be shown to ontologically depend on these fundamental structures, thus endowing the latter with concreteness in a `top-down' manner \citep[283f]{hugwut13}. Finally, the only condition necessary to evade the threat of empirical coherence is that the conditions for empirical confirmation are set in place at the scales of the human scientist, not at the level of fundamental ontology. Establishing that at human scales, the world is indeed spatiotemporal to a very good approximation circumvents the menace of empirical incoherence, at least insofar it was motivated by the apparent absence of spacetime. 

In order to establish the emergence of spacetime, it would evidently not be enough to satisfy merely necessary, but jointly insufficient conditions. But the three resisting concerns listed above may guide us in what it would take to appease them and thus arguably arrive at jointly sufficient conditions. Collectively, they suggest that in order to accomplish this, we need to focus on the functions spacetime plays in structuring our ontology, distinguish its elements from abstracta, and enabling empirical confirmation, among others. Thus, spacetime should be analysed in terms of the functions it performs to other ends, rather than for its own sake. Spacetime is as spacetime does, to invoke the motif in \citet{lamwut18}, although this need not be interpreted in ontologically thick ways. In this spirit, establishing the emergence of spacetime involves showing how the fundamental may instantiate these functions or roles in favourable circumstances. This `spacetime functionalism' we have already encountered in chapter 1 asserts that once the functional roles of spacetime have been identified and the fundamental physics shown to fulfil these roles, the emergence of spacetime has been fully established with full physical salience and no work remains to be completed in this respect. 

\subsection{Spacetime functionalism in causal set theory}
\label{ssec:sptfunct}

How would one implement a functionalist programme for CST? The first step (SF1) of spacetime functionalism as articulated in \S1.6 
will require the specification of the `spatiotemporally salient' features of spacetime that need accounting for. We will not give an exhaustive and conclusive list here, but it should be clear that these will include both topological aspects such as the dimensionality of space or spacetime and something approaching a topology of (non-metrical) nearness relations, as well as geometrical, i.e.\ metrical, relations such as (spatial) distances and (temporal) durations. Nearness and metric relations are both involved in spacetime's central role of (relative) spatiotemporal localization. In the second step prescribed by spacetime functionalism, (SF2), it must be shown how causets can fulfil these roles. This involves precisely the kind of work presented in \S2.3 
in the previous chapter, showing how the dimensionality (\S2.3.3
), spatial topologies (\S2.3.4
), and distance relations (\S2.3.5
) can emerge from the fundamental causet if the fundamental degrees of freedom come together in the right way. There, we used this work to show that space in itself cannot be usefully thought to be an inherent part or substructure of causets but that in order to recover the salient features of space, we must generally take into account the entire causet. Here, we want to recall these reconstructions because they exemplify how the second step prescribed by spacetime functionalism will look like. To repeat, such reconstructions will not be required for each and every property of relativistic spacetimes: spacetime functionalism does not demand the derivation of full relativistic spacetime, lock, stock, and barrel. 

What precisely the functions are which must be recovered from the fundamental ontology will be strongly confined by its epistemology: the crucial spatiotemporal functions to be recovered connect to ways in which we can come to know of these structures, at least in principle, and to what it takes, more generally, for the world to manifest itself to us as it does. There is a reasonable and important debate to be had what these functions are. But the constraints imposed by the epistemology arguably makes the debate more controlled than in the philosophy of mind, where the debate turns on the status of qualia. The project, then, is to explain how the structures postulated in QG play these functions crucial to the epistemology and the phenomenology, and not to articulate what the essence of the things performing these roles is. 

Although the two steps to the functionalist emergence of spacetime are sketched but in outline here, and most work remains to be done by researchers in the program, we claim that the approach offers a full and satisfactory resolution of the disappearance of salient features of spacetime in what appears to be structures rather different from what we are used to in GR. We freely admit that the recovery of empirically vital functions of spacetime from causets is far from accomplished. What we reject is the misguided insistence that even once all relevant spacetime roles are demonstrated to be fulfilled by the fundamental degrees of freedom, it would furthermore be necessary to identify the nature of spacetime in the fundamental structure, which must therefore be, by necessity, of an ultimately spatiotemporal character. Quite the opposite: we see no in principle obstacle that the fundamental structure can be wildly non-spatiotemporal---as long as it is demonstrated that it adequately plays the requisite spacetime roles at macroscopic scales. On a functionalist understanding, spacetime just is what it contributes to the overall scientific account of the world as it manifests itself to us in its myriad ways. Obviously, a quantum theory of gravity and the ontology it posits will have to assume its role in the wider context of this account. At the end of the day, it will earn its keep by lifting its share of the weight in this bigger project.

\section{Causal set theory and relativistic becoming}
\label{sec:becoming}

Now that we have sketched CST, illustrated how spacetime disappears in it, and sketched how it re-emerges, let us finish the discussion by addressing two points of considerable philosophical interest which come up in CST and in its conception of the emergence of spacetime: the possibility of a relativistically invariant passage of time or becoming, and the appearance of highly non-local behavior in this classical, relativistic, but discrete theory. We save the second point for the next section (\S\ref{sec:lorentz}) and turn to the first issue, the metaphysics of time based on CST.\footnote{This section is based on work previously published in \citet{wutcal15}.} Both points aptly illustrate the philosophical and conceptual issues that arise in developing a new quantum theory of gravity, and how spacetime emerges from it. The emergence of spacetime in quantum gravity is inextricably entangled with deep, and unavoidable, philosophical questions.

\subsection{The basic dilemma of relativistic becoming}
\label{ssec:dilemma}

Contemporary physics is notoriously hostile to a what philosopher have come to call `A-theoretic' metaphysics of time, i.e., a metaphysics of time which fundamentally includes an element of becoming or dynamical passage, an aspect of time captured in our language by the use of tenses.\footnote{As opposed to a `B-theoretic' metaphysics of time, which does not fundamentally include such elements of becoming, but instead seeks to explain them as emergent or illusory phenomena.} An important example of an A-theoretic metaphysics is \textit{presentism}, according to which only present objects and events exist, but in a dynamically updated way, such that we arrive at a metaphysics of a dynamical succession of `nows'. Another example is the \textit{growing block view}, according to which only past and present objects and events exist such that the sum total of existence continually grows by including ever more slices of existence. 

CST promises to reverse that verdict against A-theories: its advocates have argued that their framework is consistent with a fundamental notion of `becoming'. Relevantly, the dynamical `growth' of causets we introduced in \S\ref{ssec:dynamics}, this `birthing' process of new elements is said to unfold in a `generally covariant' manner and hence in a way that is perfectly compatible with relativity.  Here is \citet{Sorkin2006} advertising the philosophical pay-off: 
\begin{quote}
One often hears that the principle of general covariance [...] forces us to abandon `becoming' [...]. To this claim, the CSG dynamics provides a counterexample. It refutes the claim because it offers us an active process of growth in which `things really happen', but at the same time it honors general covariance. In doing so, it shows how the `Now' might be restored to physics without paying the price of a return to the absolute simultaneity of pre-relativistic days. 
\end{quote}
The claim is that CST as augmented with a dynamics such as classical sequential growth (CSG) dynamics, rescues temporal becoming and our intuitive notion of time from relativity. The claim is routinely made in the pertinent physics literature, and has found its way into popular science magazines.\footnote{Cf.\ e.g.\ \citet[38]{dowker2003}. For other important expressions of the claim, see \citet{Sorkin2007,dowker14,dowker20}.} One might not believe that our intuitive notion of time needs or deserves rescuing, but there is no denying that if this claim is correct it would have significant consequences for the philosophy of time. Specifically, it may underwrite a `growing block' model of the metaphysics of time, as John \citet{Earman2008} has speculated. 

As argued in \citet{wutcal15}, we believe that the introduction of CST and its dynamics does not ultimately change the fundamental dilemma any fan of becoming or passage confronts when facing relativistic physics, even though some novel aspects arise, mainly due to the discreteness of causets. The dilemma is the following: any metaphysics of time including a fundamental sense of becoming or passage either answers to their fan's explanatory demands or is compatible with relativistic physics, \textit{but not both}.\footnote{For earlier articulations of the same dillemma, see \citet{cal00} and \citet{wut13}.} 

To illustrate this with an example, suppose one thought, as does the presentist, that in order for there to be becoming or passage, there needs to be a (dynamically changing) present, a `now', identified in the fundamental structures of the world. In the context of special relativity, with its relativity of simultaneity, one might  introduce a foliation of spacetime into spacelike hypersurfaces totally ordered by `time'. Presumably, that would answer to the presentist's notion of a (spatially extended) present and of becoming, but at the price of introducing structure not invariant under automorphisms of Minkowski spacetime and hence arguably violating special relativity. Conversely, the present can be identified with invariant structures such as a single event or the surface of an event's past lightcone, and successive presents as a set of  events on a worldline or as a set of past lightcones totally ordered by inclusion, respectively, but such structures will have radically different properties from those ordinarily attributed to the present by those seeking to save it (see \citet{cal00} and \citet{wut13}). 

Returning to CST, a natural transposition of the idea of a foliation of spacetime into spacelike hypersurfaces is to partition a causet into maximal antichains, as considered in \S2.3, 
and particularly in \S2.3.2. 
Although such a partition always exists (unlike a complete foliation for relativistic spacetimes), it hardly answer to the explanatory demands of an advocate of becoming, as \citet[\S3]{wutcal15} argue in more detail. They conclude that, at least at the kinematical level, CST embraces the dilemma and in fact makes it more rigorous. But that may not be all that surprising; after all, the heart of the idea that CST rescues becoming involves taking sequential growth seriously: becoming is embodied in the `birthing' of new elements, and so in the theory's dynamics. 

Although we are interested in becoming, we should immediately remark that sequential growth is certainly compatible with a tenseless or block picture of time. In mathematics a stochastic process is defined as a triad of a sample space, a sigma algebra on that space, and a probability measure whose domain is the sigma algebra.  Transition probabilities are viewed merely as the materials from which this triad is built.  In the case at hand, the sample space is the set $\Omega = \Omega(\infty)$ of past-finite and future-infinite labeled causets that have been `run to infinity'. The `dynamics' is given by the probability measure constructed from the transition probabilities; for details, see \citet{Brightwell2003}.  On this picture, the theory consists simply of a space of complete histories with a probability measure over them.\footnote{This interpretation corresponds to Huggett's first option \citeyearpar[16]{hug14}, which is fully B-theoretic. When we consider `taking growth seriously', we mean to essentially follow the second route he offers: augmenting the causal structure with an additional, but gauge-invariant, dynamics.}

\subsection{Taking growth seriously}
\label{ssec:growth}

However, let's take the growth seriously.  There are different extents to which this can be done. At a more modest level, and consistent with explicit pronouncements by advocates of causet becoming, we can articulate a localized, observer-dependent form of becoming. Here, the idea is that becoming occurs not in an objective, global manner, but instead with respect to an observer situated within the world that becomes. The only facts of the matter concerning becoming are local, and are experienced by individual observers as they inch toward the future. In Sorkin's words, which are worth quoting in full, 
\begin{quote}
[o]ur `now' is (approximately) local and if we ask whether a distant event spacelike to us has or has not happened yet, this question lacks intuitive sense. But the `opponents of becoming' seem not to content themselves with the experience of a `situated observer'. They want to imagine themselves as a `super observer', who would take in all of existence at a glance. The supposition of such an observer {\em would} lead to a distinguished `slicing' of the causet, contradicting the principle that such a slicing lacks objective meaning (`covariance'). \citeyearpar[158]{Sorkin2007}
\end{quote}

According to Sorkin, instead of ``super observers'', we have an ``asynchronous multiplicity of `nows' ''. It seems fairly straightforward that a perfectly analogous kind of becoming can be had in the context of Minkowski spacetime. Indeed, `past lightcone becoming', in the sense of \citet{Stein1991}, and `worldline becoming', as articulated by \citet{CliftonHogarth}, both satisfy the bill.\footnote{Cf.\ also \citet{Arageorgis12} who makes a similar point.} Furthermore, past lightcone becoming and worldline becoming are also available in general-relativistic spacetimes, as they do not depend on the spacetime admitting a foliation. 

Although Sorkin himself remains uncommitted concerning whether the analogy holds, Fay \citet{dowker14,dowker20} rejects it, arguing that `asynchronous becoming' is not compatible with GR, but only with a dynamics like the one provided by the classical sequential growth. She discusses `hypersurface becoming' in GR---which does of course depend on the spacetime admitting a foliation---and rejects it for its obvious violation of general covariance. She is clear that hypersurface becoming is but one way to implement becoming in the context of GR, but does unfortunately not discuss alternatives. In particular, she does not discuss worldline becoming or past lightcone becoming, which both seem much more promising analogues of the asynchronous becoming identified by Sorkin in CST. She maintains that what is needed for there to be becoming is not the mere \textit{existence} of events, but a process she term terms their \textit{occurrence} \citeyearpar[22]{dowker14}. She claims that spacetime events do not `occur' in GR in this special sense, and so there is no genuine form of becoming possible in GR. Against this, we note firstly that (an important subsector of) general relativity certainly can be described in a `dynamical' manner via its many `3+1' formulations.\footnote{Cf.\ \citet[Ch.~10]{Wald1984}.}  To make her objection, Dowker would first need to elaborate the reasons why a 3+1 dynamics does not provide the `occurrence' she desires. Of course, we admit that our retort here lacks full generality. 

Furthermore, we note here a possible tension.  If occurrence is simply a label for some events from the perspective of other events, then there is no problem---but then we note that such labels can be given consistently in GR too.  But if occurrence implies something metaphysically meaty, such as existence or determinateness---then there is a possible tension between occurrence and the local becoming envisioned by Sorkin and Dowker.  If spacetime events that are spacelike related do not exist for each other, for instance, then that is a radical fragmentation of reality.\footnote{In Pooley's view \citeyearpar[358n]{Pooley13}, dynamical classical sequential growth should best be interpreted as a ``non-standard $A$ Theory'' in the sense of \citet{Fine2005}, i.e., as giving up ``the idea that there are absolute facts of the matter about the way the world is.'' \citeyearpar[334]{Pooley13}}  Not only would that be a high cost to introduce becoming, but it is also one that, again, could be introduced in GR. 

Our present interest is to determine whether a more ambitious, objective, global, observer-independent form of becoming is compatible with classical-sequential-growth-cum-dynamics in a way that does not violate the strictures of relativity. In other words, does Sorkin's assertion in the last sentence of the indented quote above hold up to scrutiny? We will argue that it does not and that there is a weak sense in which a fully objective kind of becoming with relativistic credentials can be had. 

The central problem with taking the primitive growth of classical sequential growth dynamics as vindicating becoming is that this dynamics uniformly treats the labeling time as `fictitious'.  It is only `real' to the extent to which it respects the causal order (in the sense of axiom \ref{ax:temp}). A choice of label would be tantamount to picking a time coordinate $x_0$ in a relativistic spacetime.  Any dynamics distinguishing a particular label order will be non-relativistic.  Not wanting the dynamics to distinguish a particular label (`coordinatization'), the authors impose {\em discrete general covariance}, i.e., axiom \ref{ax:gencov} on the dynamics.  This is a form of label invariance.  As stated above, the idea is that the probability of any particular causet arising should be independent of the path to get to that causet. In fact, returning to figure \ref{fig:transpercolation}, there could be no physical fact of the matter whether the 3-element causet in the middle of the top row grew via a stage consisting in a 2-chain or a 2-antichain (see also figure 2 in \citealt{wutcal15}). In the 3-element causet, the event with another event in its past (labelled `$a$' in figure \ref{fig:transpercolation}) and the causally isolated event (labelled `$b$') are spacelike related. Consequently, there is not fact of the matter which of the two came into being first and which one second. To say which one happened `first' is to invoke non-relativistic concepts.  It is therefore hard to understand how there can be growth happening in time. 

Seeing the difficulty here, John \citet{Earman2008} suggests a kind of philosophical addition to causets, one where we imagine that `actuality' does take one path or another.  With such a hidden variable moving up the causet, we do regain a notion of becoming.  But as Aristidis \citet{Arageorgis12} rightly points out, such a move really flies in the face of the normal interpretation of these labels as pure gauge.\footnote{Cf.\ also \citet[859f]{Butterfield2007}.} The natural suggestion, espoused by (almost all?) philosophers of physics, is then that the above non-dynamic interpretation in terms of a block universe is best because it does not ask us to imagine that one event came first.

Perhaps the sensible reaction to this problem is to abandon the hope that the classical sequential growth dynamics does produce a novel sense of becoming.  Still, we are tempted to press on.  The intuition motivating us is as follows.  True, the dynamics is written in terms of a choice of label, but we know that a consistent gauge invariant dynamics exists `beneath' this dynamics.  In fact, rewriting the theory in terms of a probability measure space, as indicated above, one can quotient out under relabellings to arrive at a label-invariant measure space (for construction and details, see \citealt{Brightwell2003}).  And one thing that we know is gauge invariant is the number of elements in any causet.  Focusing just on these and ignoring any labeling, we do have transitions from $\mathcal{C}_n$ to $\mathcal{C}_{n+1}$ and so on.  There is gauge-invariant growth.

The problem is that we are generally prohibited from saying exactly what elements exist at any stage of growth.  Take the case of the spacelike related events above.  The world grows from $\mathcal{C}_1$ to $\mathcal{C}_2$ to $\mathcal{C}_3$. That's gauge invariant.  We just cannot say---not due to ignorance, but because there is no fact of the matter---whether $\mathcal{C}_2$ is the 2-chain or the 2-antichain.  causet reality does not contain this information. There simply is no determinate fact as to whether $\mathcal{C}_2$ in the 2-chain or the 2-antichain; but there is a determinate fact that it contains one of them. If it is coherent, therefore, to speak of a causet having a certain number of elements but without saying what those elements are, then classical sequential growth dynamics does permit a new kind of---admittedly radical and bizarre---temporal becoming. 

Whether this notion of becoming is coherent depends on the identity conditions one has for events.  If to be an event, one has to be a particular type of event with a certain character, then perhaps the idea is not coherent. After all, what is the $\mathcal{C}_2$ world like?  It does not have the 2-chain {\em and} the 2-antichain in it (that's $\mathcal{C}_3$), nor does it have neither the 2-chain nor the 2-antichain in it (that's $\mathcal{C}_1$).  The world determinately has the 2-chain or the 2-antichain in it, but it does not have determinately the 2-chain or determinately the 2-antichain.  `Determinately' cannot penetrate inside the disjunction.  Notice that this feature is a hallmark of vagueness or of metaphysical indeterminacy more generally. Without going into any details of the vast literature on vagueness, let us note that there is a lively dispute over whether there can be ontological vagueness. The causet program, interpreted as we have here, supplies a possible model of a world that is ontologically vague.  Further discussion of this model seems to us worthwhile.

We would like to point out that Ted \citet{Sider03} has supplied arguments that existence cannot be vague. That existence cannot be vague or indeterminate was a central assumption of his argument to four-dimensionalism in his \citeyearpar{Sider01}. In fact, he asserts \citeyearpar[135]{Sider01} that anyone who accepts the premise that existence cannot be vague is committed to four-dimensionalism, the thesis that objects persist by having temporal parts. To the extent to which many advocates of becoming reject four-dimensionalism anyway, they would thus be open to embrace ontological indeterminacy even if Sider's arguments of 2001 and 2003 were successful. And they may well not be: one of them, for instance, is based on the claim that it cannot be vague how many things there are in a finite world \citeyearpar[136f]{Sider01}. Obviously, a defender of observer-independent becoming in CST may agree that it is at no moment vague how many events there exist, but nevertheless disagree that existence cannot be vague. Thus, we may have ontological indeterminacy without vagueness in the cardinality of the (finite) set of all existing objects. 

\subsection{Some bizarre consequences}
\label{ssec:bizarre}

\begin{figure}[ht]
\centering
\epsfig{figure=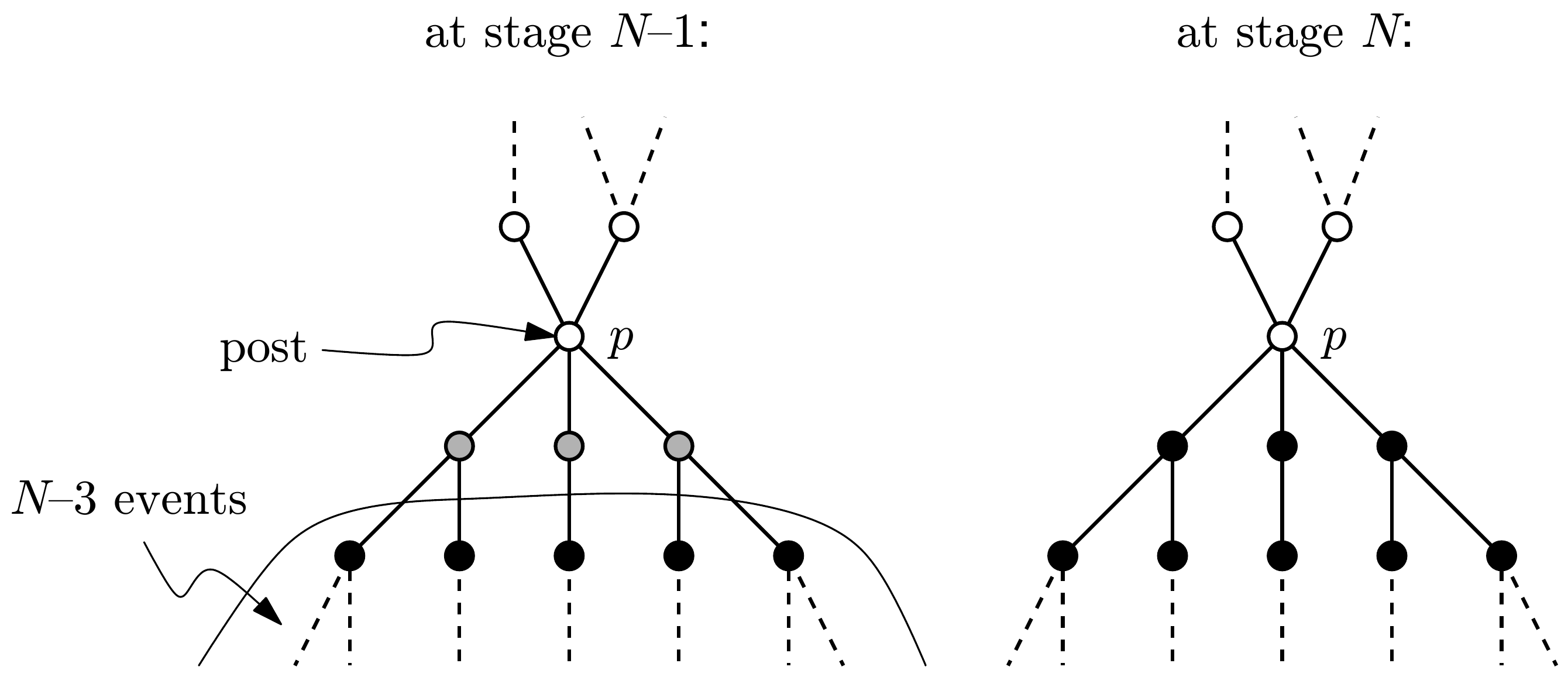,width=0.78\linewidth}
\caption{\label{fig:postgrowth} Becoming at post $p$ (Figure from \citealt{wutcal15}).}
\end{figure}

One may be worried that on this notion of becoming in CST, no event in a future-infinite causet may ever be determinate until future infinity is reached, at which point everything snaps into determinate existence. This worry is particularly pressing as realistic causets are often taken to be future-infinite. So does any event ever get determinate at any finite stage of becoming? In general, yes. One way to see this is by way of example. As it turns out, causets based on transitive percolation in general have many `posts', where a {\em post} is an event that is comparable to every other event, i.e., an event that either is causally preceded by or causally precedes every other event in the causet. Rideout and Sorkin interpret the resulting cosmological model as one in which ``the universe cycles endlessly through phases of expansion, stasis, and contraction [...] back down to a single element.'' \citeyearpar[024002-4]{Rideout1999}\footnote{Cf.\ also \citet{Bollobas1997}.} Consider the situation as depicted in figure \ref{fig:postgrowth}.
There is a post, $p$, such that $N$ events causally precede $p$, while all the others---potentially infinitely many---are causally preceded by $p$. At stage $N-1$, shown on the left, there exist $N-1$ events. At this stage, all the `ancestors' of $p$ except those three events which immediately precede $p$, shown in black, must have determinately come to be. Of the three immediate predecessors, shown in grey to indicate their indeterminate status, two must exist; however, it is indeterminate which two of the three exist. At the prior stage $N-2$, the grey set of events existing indeterminately would have extended one `generation' further back, as it could be that two comparable events are the last ones to come to be before the post becomes. At the next stage, stage $N$, $N$ events exist and it is determinate that all ancestors of $p$ exist. There is no ontological indeterminacy at this stage. Event $p$ has not yet come to be at either stage and is thus shown in white. At stage $N+1$, not shown in figure \ref{fig:postgrowth}, event $p$ determinately comes into existence. At stage $N+2$, one of the two immediate successor to $p$ exists, but it is indeterminate which one. And so on. 

One may object that this interpretation of the dynamics of a future-infinite causet presupposes a given final state toward which the causet evolves, and thus involve a teleological element. Even though everything in the preceding paragraph is true under the supposition that the final causet is the one represented in figure \ref{fig:postgrowth}, the objection goes, at stage $N$ it is not yet determined {\em that} $p$ is a post, as there could have been other events spacelike-related to $p$. Given that it is thus indeterminate whether $p$ is indeed a post, and since this is the case for all events at finite stages, no events can thus snap into determinate existence at any finite stage of the dynamical growth process. 

First, it should be noted that even if this objection succeeds, it is still the case that it is objectively and determinately the case that at each stage, one event comes into being and that thus the cardinality of the sum total of existence grows. Although the ontological indeterminacy remains maximal, there is a weak sense in which there is objective, observer-independent becoming. Second, if the causet does indeed not `tend' to some particular future-infinite causet, then all existence would always be altogether indeterminate (except for the cardinality). There would be no fact of the matter, ever, i.e., at any finite stage, of how the future will be, or indeed of how anything ever is. If this is the right way to think about the metaphysics of the dynamics of CST, we are left with a wildly indeterminate picture. Third, it should be noted that the mathematics of the dynamics is only well-defined in the infinite limit; in particular, for there to be a well-defined probability measure on $\Omega$, we must take $\Omega = \Omega(\infty)$ (\citealt[160n]{Sorkin2007}; \citealt[\S3]{Arageorgis12}), which can be interpreted to mean that the future-infinite `end state' is metaphysically prior to the stochastic dynamics that grows the causet to that `state'. 

We close this section with a discussion of some of the strange features of this metaphysics.  First, note that many philosophers, from Aristotle to today, have thought that the future is indeterminate (see, e.g., \citealt{OhrstromHasle11} and references therein).  According to some versions of this view, it's determinately true that tomorrow's coin flip will result in either a head or a tail, but it is not determinate yet which result obtains.  Vagueness infects the future.  We note that the above causet vagueness is quite similar, but with one big difference:  on the causet picture, the past too can be indeterminate!  In our toy causet, it is not true at $\mathcal{C}_3$ that $\mathcal{C}_2$ determinately is one way rather than the other.  

Second, note that as a causet grows, events that were once spacelike to the causet might acquire timelike links to future events.  If we regard the growth of a new timelike link to a spacelike event as making the spacelike event determinate, modulo the above type of vagueness, then this is a way future becoming can make events past.  That is, there is a literal sense in which one can say that `the past isn't what it used to be'. Having said that, there is a relativistic analogue in the growth of past lightcones, which also come to include formerly spacelike-related events as one moves `up' along a worldline. 

Finally, although we don't have space to discuss it here, note that despite appearances transitive percolation is perfectly time reversal invariant.  This allows the construction of an even more exotic temporal metaphysics.  If we relax the assumption that events can only be born to the future of existing events, then it is possible to have percolation---and hence becoming---going both to the future and past.  Choose a here-now as the original point.  Then it is possible to modify the theory so that the world becomes in both directions, future and past. Of course, similarly, we could have a causet that is future-finite and only grows into the past, and thus is past-infinite.

In sum, then, does CST rescue temporal becoming? At the kinematical level (not much discussed here, but see \citealt{wutcal15}), CST does offer new twists in dealing with time and relativity, but the basic contours of the relativistic challenge remains.  Serious constraints also threaten becoming if we take the time in CST's dynamics seriously too.  Here, however, if one is open to the costs of a sufficiently radical metaphysics, we maintain that there is a novel and exotic type of objective, observer-independent temporal becoming possible. It should be noted, again, that all this remains purely classical and so subject to potentially radical change in a future quantum theory of causets. So far, it is really just a classical, discrete, and dynamically stochastic theory. The discreteness of its structure has another unusual consequence: a classical form of non-locality.

\section{Lorentz symmetry and non-locality}
\label{sec:lorentz}

As we have discussed in chapter 2
, the fundamental discreteness of causets has certain conceptual and technical advantages, as it promises to eliminate the nasty ultraviolet divergencies of quantum field theory (QFT) and the singularities of GR. However, it seems to also come at a price: it is incompatible with the demand for Lorentz symmetry and the expectation of a reasonably local physics. It seems as if one can have any two of these three features---discreteness, Lorentz symmetry, and locality---in a theory, but not all three, as was already noted by \citet{Moore1988} within half a year of the publication of the founding document of the causet program \citep{bomeal87}. There is of course a sense in which no discrete structure can be Lorentz invariant, as Lorentz transformations are only defined for continuum spacetimes. However, even is the fundamental structure is not Lorentz invariant, one would not want to give up Lorentz symmetry at the emergent level, at least to an extremely close approximation. This places a very strong constraint on both the spacetime structures and the dynamics of matter that can emerge from causets. As discreteness is built into the DNA of the causet program, it seems as if CST is committed to a form of non-locality, at least at the fundamental level. But our world seems manifestly local at all observed scales---apart from quantum non-locality of course---, and so local physics better emerge from the fundamental physics of causets.\footnote{The best references explaining this resulting non-locality in the context of CST are \citet{sor09} and \citet{dowker2011}.}

\subsection{Lorentz symmetry and discreteness entail non-locality}
\label{ssec:incompatible}

How should we understand the claim that the demand for Lorentz symmetry and the discreteness of causets entail a form of non-locality of the physics? The discreteness of causets in built in from the start---in axiom 2.1
. Lorentz symmetry is the demand common in relativistic physics that the dynamics be invariant under Lorentz transformations. In a special-relativistic theory, such as the standard model of elementary particle physics, this demand is encoded in the invariance of the theory's action under Lorentz transformations and again in the spacetime symmetries of Minkowski spacetime. In GR, spacetimes in general no longer have global symmetries like Lorentz symmetry, but Lorentz symmetry still holds `locally', i.e., at sufficiently small scales. Consequently, causets must give rise to spacetimes which are at least locally Lorentz symmetric. In this sense, Lorentz symmetry must at least hold at some mesoscale: it may be violated at the fundamental level, and it may not hold as a global symmetry of emergent spacetime, but there must be a robust range of scale where it holds to an extremely good approximation. No violation has ever been detected,\footnote{In the sense that the bounds for violating Lorentz symmetry are extremely tight \citep[\S2.1.2]{Will2014}.} and so the demand is on empirically rather secure grounds.

In order to recognize the non-locality of a Lorentz-invariant, discrete structure, let us return to the standard way of conceiving of the relation between causets and spacetime as depicted in figure \ref{fig:emergingspt3}, prototypically captured by a Poisson sprinkling of elements into Minkowski spacetime (as in \S\ref{ssec:manifold}). The sprinkling must be random, or erratic, in that it cannot be too regular. If it had an even just somewhat regular lattice structure, then it would involve preferred directions in that a distribution of elements in Minkowski spacetime will look different from its Lorentz-transformed cousins. Thus, the demand for Lorentz symmetry translates into a requirement that the sprinkling be random \citep{bomeal09}. In a Poisson sprinkling for instance, the probability of sprinkling $n$ elements into a spacetime region of spacetime volume $V$ depends only on the density $\rho$ of the sprinkling and $V$:
\begin{equation}
P(n) = \frac{(\rho V)^n e^{-\rho V}}{n!}.
\end{equation}
Since $\rho$ is fixed, this probability is manifestly invariant under all transformations preserving spacetime volume, such as Lorentz transformations \citep[\S12.2.1]{Henson2012}. 

\begin{figure}[ht]
\centering
\epsfig{figure=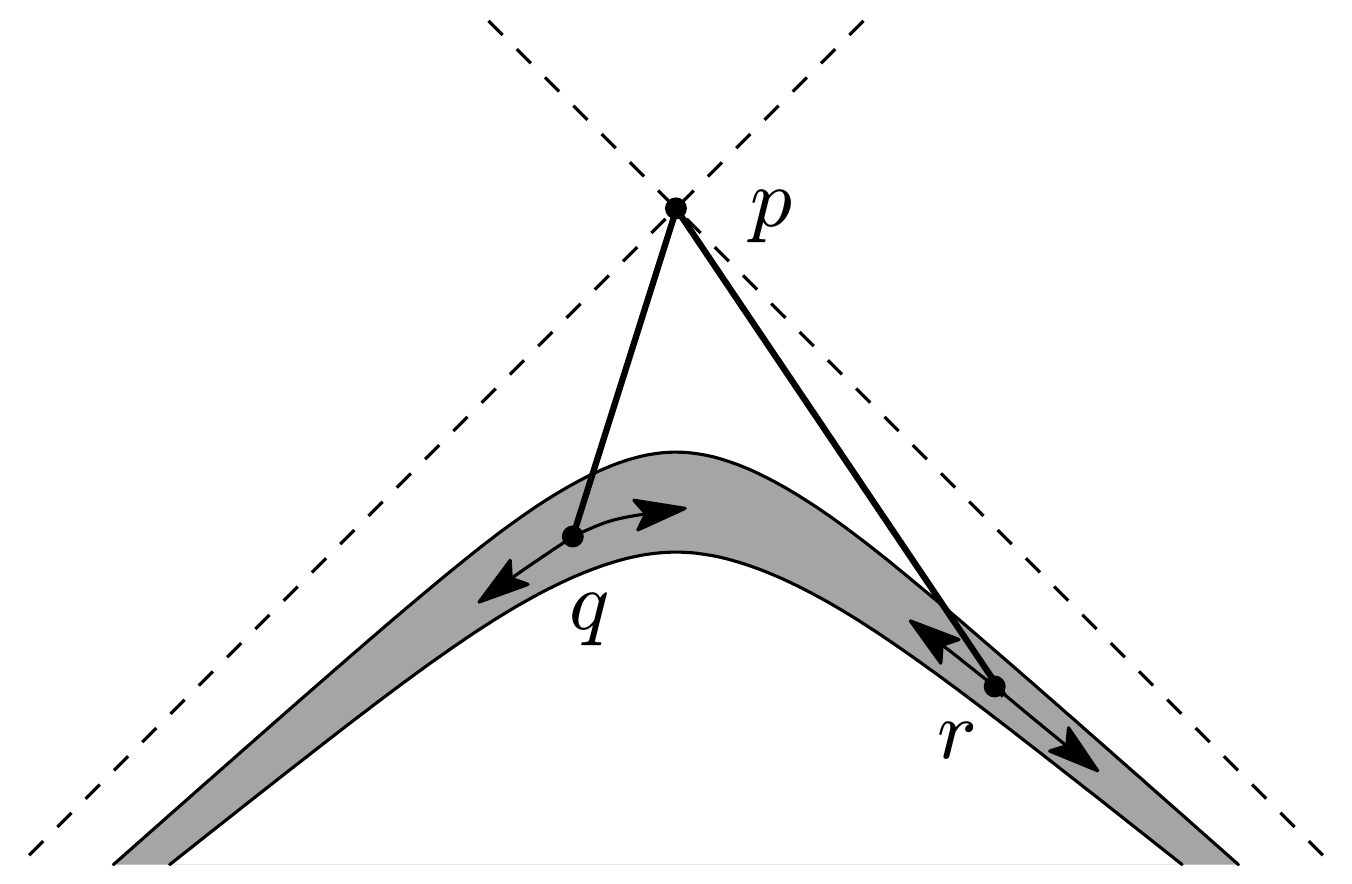,width=0.42\linewidth}
\caption{\label{fig:nonlocal} There are infinitely many events in the thickened past hyperboloid of event $p$.}
\end{figure}

The density $\rho$ of the sprinkling is a fundamental constant of the theory, but is naturally set at something like the Planck scale. As basically noted by \citet{Moore1988}, the spacetime volume of a hyperboloid shell of any finite thickness (defined spatiotemporally by an invariant interval $\Delta s$) at a given spatiotemporal distance in the past from any event $p$ is infinite. The greyed out region in figure \ref{fig:nonlocal} represents such a hyperboloid shell to the past of an event $p$. Thus, whatever the fixed, constant density $\rho$ is, the number of elements in the hyperboloid shell diverges. 
If the hyperboloid shell is, for example, at a Planck interval to the past of $p$, this means that $p$ will have infinitely many predecessor elements a Planck distance in its past. In fact, whatever the spacetime interval chosen as the scale, and whatever the (finite) thickness of the shell, this will invariably be the case. Moreover, the spacetime volume of the region in $p$'s past lightcone which consists of points \textit{less} than the chosen interval in the past (i.e., the past lightcone of $p$ minus the hyperboloid shell and \textit{its} past) is also infinite and so contains infinitely many sprinkled elements. Thus, within any fixed spacetime interval to the past of an element of a causet being approximated by a Minkowski spacetime, there are necessarily an infinite number of elements. In this sense, there is no nearest neighbour to the past of any given event in a sprinkling of Minkowski spacetime.\footnote{`Nearest' as measured in terms of the invariant spacetime interval $\Delta s$.} For any event sprinkled close to $p$ in $p$'s past, there is always another event sprinkled to the past of $p$, which is even closer to $p$ \citep{bomeal09}. The same holds for immediate neighbours to the future.\footnote{We are not interested in spacelike relations between events, even though the same applies \textit{mutatis mutandis} for spacelike related neighbours.} This is a form of non-locality in that in any inertial frame, events arbitrarily far away spatially from $p$ (in that frame) will be arbitrarily close to $p$ in terms of the invariant spacetime interval. 

Inevitably, it thus seems as if any element of any causet being approximated by Minkowski spacetime must have an infinite number of `immediate' predecessors and an infinite number of `immediate' successors.\footnote{What is an `immediate' or `nearest' neighbour? In the fundamental theory, it is defined as being directly related. In Minkowski spacetime, the sprinkled events are `immediate' or `nearest' neighbours if the Alexandrov interval defined by the two events does not contain another event. The Alexandrov interval is defined as the non-empty intersection of the causal future of one event and the causal past of the other.} As \citet[655]{Moore1988} concludes, it seems as if discrete `spacetime' ``has the nasty property that every point is influenced by an infinity of `nearest neighbors' which, in a given frame, are arbitrarily far back in time.'' 

In their reply, \citet{bomeal88} offer two lines of defence, both of which are sound in our view. First, they point out that this is really the natural discrete analogue of the fact that an event in a Lorentz manifold has neighbourhoods which converge to its lightcones as well, and so \textit{should} be reflected by causets. In this sense, in a continuum spacetime, events spatially arbitrarily far away from $p$ in a given frame are also arbitrarily close to $p$ as measured by the spacetime interval. Perhaps the non-locality inherent in causets is really not all that troubling after all. Second, they point out that (general-)relativistic spacetimes are only locally Lorentz invariant and that in important cases such as the Friedmann-Lema{\^\i}tre-Robertson-Walker spacetimes, the past lightcones of each event has a finite volume and so a sprinkling would suggest only a finite number of events to the past in a discrete analogue. In fact, they assert, this will be so in the general situation using a `sum-over-histories' approach, which they prefer. 

This latter point deserves some unpacking. In general, Lorentz invariance will not turn out to be a global symmetry of emergent spacetime. In this sense, spacetime curvature will limit Lorentz symmetry, restricting the number of nearest neighbours of an element in a underlying discrete structure to a finite number. However, unless the spacetime curvature is rather large, i.e., the radius of curvature comparable to the Planck length, the number of nearest neighbours of any element will still be very large. Hence, fundamental physics according to CST cannot be `local' in that the physics at an element of the causet depends only on that of a neatly confined, finite (and somewhat `small') set of elements. 

A remark before we proceed. This non-locality becomes apparent only once we consider embeddings of causets into Minkowski spacetime and countenance the symmetries of that spacetime. One might thus be tempted to think that the non-locality is somehow an artefact of this conception of the relation between causets and spacetimes, an artefact which might disappear if we replace the Poisson sprinkling of events into Minkowski spacetime with a appropriate form of relating the two. But this would be too quick: the argument shows, quite generally, that a causet whose elements do not have a very large number of immediate neighbours cannot possibly give rise to a spacetime with at least approximate Lorentz symmetry. Thus, the high valency of its elements should be expected of any physically realistic causet.

What to make of this non-locality? As \citet{Benincasa2010} explain, this non-locality is both a blessing and a curse vis-\`a-vis two central challenges of any approach to quantum gravity trading in fundamentally discrete structures. The first goal of quantum gravity is to explain how the dynamics manages to drive these discrete kinematic structures toward those that are well-approximated by Lorentzian manifolds. In this respect, they state, the non-locality is a blessing, as a purely local dynamics could not account for the manifold-likeness of causets. The reason for this failure is that any local action would have to be expressed as a sum over the contributions from each element of the discrete structure, and thus could not grow faster than linearly with the number $n$ of elements. Assuming that it is the dynamics which tames the KR catastrophe (see \S\ref{ssec:setup} and \S\ref{ssec:dynamics} above), and since the number of causets with $n$ elements grows as $e^{n^2/4}$, their contributions to the partition function risk being outcrowded by those of non-manifold-like terms. Although this is of course not a conclusive argument, Benincasa and Dowker thus do not expect a local dynamics to be able to account for the manifold-likeness and thus not be a solution in the sense of \S\ref{ssec:dynamics}. 

The second challenge of a discrete approach to quantum gravity according to \citet{Benincasa2010} is to explain why the geometry of the emergent spacetime should be a solution to the Einstein field equation. With regard to this challenge, the non-locality turns out to be more curse than blessing, as local dynamics and the demand for general covariance ``pretty much guarantee'' \citep[2]{Benincasa2010} that the Einstein equation be satisfied. In the absence of a local dynamics, there is no such guarantee, and it remains an open question whether CST or a similarly non-local discrete approach to quantum gravity can explain how the geometry of emergent spacetime conforms to GR. They fear that ``if causets were incorrigibly nonlocal, this would be fatal'' (ibid.) for the project. But fortunately, there is hope.

\subsection{Implications: hierarchy of scales and phenomenological signatures}

This hope will be fulfilled if there exists an intermediate scale at which effective physics is approximately local. In the previous subsection, we have seen that the physics of CST must be non-local at the fundamental level. Due to the presence of spacetime curvature, the global structure of spacetime will not be Minkowskian, and hence there is no requirement that the physics be Lorentz-invariant at cosmological scales, and it may well be local. However, to repeat, physics must definitely be Lorentz-invariant across a wide range between the fundamental and the cosmological scales. Thus, whatever the fundamental physics, and in particular the fundamental dynamics, it must give rise to quasi-local physics, which is Lorentz-invariant to an extremely close approximation. What exacerbates the situation, as argues \citet[28]{sor09}, is that as the non-local couplings by far outnumber the local ones, the non-locality cannot be entirely restricted to the fundamental level. 

However, as Sorkin goes on to explain, there is promising evidence that the emerging physics is reasonably local. He considers (in \S3.1) the limited, but paradigmatic example of a massless, classical scalar field $\phi$ in two dimensions with the standard equation of motion, i.e., $\Box \phi =0$, where `$\Box$' is the d'Alembertian operator. It is the significant achievement of this paper to propose a `discretized' version of a two-dimensional d'Alembertian in terms native to the fundamental causet (``fully intrinsic'', in Sorkin's words \citeyear[33]{sor09}).\footnote{Assuming that $\Box$ acts linearly on the field $\phi$, as Sorkin does, the d'Alembertian for a causet $\langle C, \prec\rangle$ can be expressed as a suitable matrix $B_{xy}$, where $x$ and $y$ range over the elements of the causet. Sorkin also requires that $B$ is `retarded' or `causal' in that $B_{xy} = 0$ in case $x$ is spacelike to, or precedes, $x$. Using a series of informed guesses based on methods in theoretical physics and computer simulations, \citet[33]{Sorkin2009b} proposes that (ignoring a constant scale factor) 
\[
B_{xy} = \left\{
  \begin{array}{ll}
    -\frac{1}{2} & : x=y\\
    1, -2, 1 & : x\neq y \;\mbox{for}\; n(x,y) = 0, 1, 2, \;\mbox{respectively},\\
    0 & : \mbox{otherwise},
  \end{array}
\right.
\]
where $n(x,y)$is the cardinality of the order interval $\langle y,x\rangle = \{z\in C| y \prec z \prec x\}$. This d'Alembertian is local in that the physics at events removed by more than two `ticks' has no influence. However, it is still non-local in that events which are immediately causally connected will be arbitrarily far away spatially in some frames.} This d'Alembertian is thought to exemplify the sort of tools needed to construct a `field' theory on the basis of CST, which will give rise, ultimately, to standard QFT. Sorkin's d'Alembertian includes non-local couplings at the fundamental level, which, however, will be suppressed at larger scales, or so he argues. Sorkin thus outlines a research program for developing the tools of recovering established physics from fundamentally discrete and thus highly non-local physics. \citet{Benincasa2010} add to this evidence by formulating a more general non-local operator, which approximates the local d'Alembertian for a scalar field in four-dimensional flat spacetime.\footnote{\citet{GlaserSurya13} propose an order-theoretic characterization of local neighbourhoods in manifoldlike causets, thus providing a frame for this as well as further evidence.} Their operator is effectively local as well, courtesy of the non-local contributions to magically cancel out, leaving just `local' contributions. Here, `local' refers to the frame defined by $\phi$ itself, i.e., to the frame in which $\phi$ varies `slowly'. It turns out that the operator is effectively local in this frame, with non-local contributions suppressed.\footnote{We thank Fay Dowker for private correspondence on this point.}

Although it must be clearly stated that none of this suffices for anything like a definitive verdict on the matter, these preliminary results amount to a `proof of concept', as \citet[4]{Benincasa2010} insist, in that they establish ``the mutual compatibility of Lorentz invariance, fundamental spacetime discreteness and approximate locality''. 

One upshot of this incipient research is also that the fundamental non-locality in CST cannot be completely tamed. Nor should it be: the proposed physics should ultimately have empirically detectable consequences, after all. As the theory delivers Lorentz-invariant mesoscopic physics by construction, no violation of Lorentz symmetry should be expected in the approach. Thus, the phenomenology must come from elsewhere, and the non-local discreteness seems a promising source. We will close the chapter by discussing two such potential effects: swerving particles and the cosmological constant.

\subsection{Swerving particles: putting matter on causets}
\label{ssec:swerving}

An issue wide open for CST is how causets interact with matter, e.g., with particles or non-gravitational fields. The d'Alembertian constructed in \citet{Benincasa2010} and related research efforts paint a way in which one might see fields propagate on causets. How should we put matter `on' causets even kinematically, prior to giving it the proper dynamics? In the simple case of a (real-valued) classical scalar field, one just sticks a number on each element of the causet. More precisely, a scalar field on a causet is represented by a map from the causet to the field $\mathbb{R}$ of real numbers. The `dynamics' then gives us a rule restricting the relations among these numbers. Vector fields, complex-valued fields, quantum fields---although more complicated---will intuitively be constructed along similar lines, mutatis mutandis.

What about particles? Restricting ourselves to the simpler case of point particles, a particle will presumably trace out a timelike path on a causet. In other words, its worldline will consist in a set of pairwise causally related elements of the causet, i.e., on a chain of causet elements. Any chain in a causet is a kinematically possible worldline for a point particle. Surely, we will want to restrict the physically possible trajectories to some subset of such worldlines. Thus, we also need to formulate dynamical rules which determine, at the emergent level at least, some standard of inertial or geodesic and forced motion. This is a challenging task, which remains to be solved in full generality. 

However, \citet{DowkerHensonSorkin03} have proposed a simple model of a point particle moving on a causet sprinkled into Minkowski spacetime. This allows the use of the background Minkowski spacetime as a standard for inertial or accelerated motion. In the simple model, the task is to determine the continuation of the particle's trajectory given an initial portion of that trajectory. A central assumption of the model is that the determination is quasi-local in time: the particle's past trajectory within a certain proper time to the past fully determines its future continuation. Dowker, Henson, and Sorkin then propose an approximately Markovian, Lorentz-invariant dynamical rule, which, they argue, is the best discrete analogue of geodesic motion in Minkowski spacetime. Within going into the details here,\footnote{A useful illustration can be found in figure 1 in \citet{dowker2011}. See also \citet[\S21.3.2]{hen09}, apart from \citet{DowkerHensonSorkin03} of course.} the basic idea is that the continuation which best preserves linear momentum (in the frame in which the three-momentum going `into' the event vanishes) is the dynamically chosen one. In general, the three-momentum going `forward', although minimized, will not be zero. In this sense, the particle appears to `swerve' away from the geodesic, apparently undergoing a random acceleration. Since this swerving diverges from the expectation of strictly geodesic motion, it would be in principle an empirically detectable signature of CST. 

That the effect size would be rather small compared to observationally accessible scales is just a practical problem. There are also more principled reasons to be critical of the model. First, the particles are not modelled in a credible manner: realistic particles will not be point particles, and be quantum in nature, so that the classical, deterministic rule above seems ill-fitted. Second, the model makes ineliminable use of Minkowski spacetime to deliver a standard of inertial motion. Surely, the details of the particle motion should not depend on the embedding of the causet into a spacetime, but be intrinsically defined in the causet. Third, and perhaps most damning, there is no sense in which the dynamics includes a backreaction of the matter content on the causet, as we would expect from GR. There are ways to overcome some of these difficulties (see e.g.\ \citealt{Philpott2009}), but research continues.

\subsection{The ever-present $\Lambda$}
\label{ssec:lambda}

Let us end with a brief discussion of CST's most suggestive `prediction', viz., that the cosmological constant $\Lambda$ has a small, but non-zero positive value of the order of $10^{-120}$ in natural units. The `prediction' earns its scare quotes due to its being a heuristic, and certainly defeasible argument, rather than a tight quantitative calculation. But it clearly is a \textit{prediction} in that it has been made prior to the relevant observations, and not a mere ex post facto construction of a just-so justification. And this makes it rather intriguing. The prediction is due to Rafael Sorkin and appears to date from the late 1980s, roughly a decade before the discovery of the accelerating of the universe's expansion in 1998 by Saul Perlmutter, Adam Riess and Brian Schmidt and their teams. The first brief published version of the argument can be found in the last paragraph of \citet{Sorkin1991}, a published version of the talk given a year earlier, in 1990. We here mostly follow \citet[\S7.2]{sor97}. 

The central idea is that the non-vanishing cosmological constant arises not from a form of `dark matter', but instead is just due to fluctuations in the volume $V$ of spacetime. Thus, it offers an explanatory paradigm for a small positive $\Lambda$ (or for the accelerated expansion), which differs interestingly from the standard account invoking dark matter. 

CST uses, as its only measure of `size', the number $N$ of causet elements. Earlier (\S\ref{ssec:dynamics}), we have seen that $N$ is a (in fact, \textit{the}) generally covariant quantity in classical sequential growth dynamics. Thus, $N$ is the measure of the spacetime volume $V$ of emergent spacetime. As a second ingredient, the argument uses the idea (borrowed from unimodular gravity) that the cosmological constant $\Lambda$ is in some sense conjugate to $V$. In a loose analogy to the energy-time uncertainty, with $\Lambda$ an `energy' and $V$ a `time', we have $\Delta V\Delta\Lambda \sim 1$ (in natural units). 

The third input required is the recognition that in CST, returning to the paradigm of a Lorentz-invariant sprinkling of causets into the approximating spacetime, is that although $N$ is a measure of $V$, the latter suffers from Poisson fluctuations in $N$, with a typical magnitude of $\sqrt{N}$. Consequently, for a fixed $N$ in the fundamental causet, $V$ will only be fixed up to fluctuations of magnitude $\sqrt{N} \sim \sqrt{V}$. Thus, the resulting uncertainty of $V$ is given as $\Delta V \sim \sqrt{V}$. Putting this together, we obtain for the uncertainty of $\Lambda$:
\begin{equation}
\Delta\Lambda \sim \frac{1}{\sqrt{V}}.
\end{equation}
Assuming that $\Lambda$ fluctuates about zero (by whatever mechanism), then what we observe today is only this fluctuation, i.e., $\Lambda \sim V^{-1/2}$. 
Further assuming that $V$ is a four-dimensional `Hubble sphere', we replace $V$ by the fourth power of the Hubble radius $H^{-1}$, where $H$ is the Hubble constant, and thus come to expect
\begin{equation}\label{eq:hubble}
\Lambda \sim H^2,
\end{equation}
which turns out to be of the order of the $10^{-120}$ in natural units mentioned above for present-day values. This is remarkably close to values derived from astronomical observations. 

A few remarks in closing. First, it should be noted that (\ref{eq:hubble}) holds for any cosmic epoch. Unlike for other approaches where $\Lambda$ can vanish during some epochs, it is thus `ever-present' here. Second, given the immense advances in observational cosmology, the heuristic argument presented above, although suggestive, will need to be supplemented by physically more rigorous and quantitatively more precise models based on CST in order to fulfil the explanatory promise issued by the above argument. \citet{everpresent} offer the beginning of a more sophisticated analysis; other models exist. Third, just as we should be used to by now, both the heuristic argument reproduced here and the more sophisticated model in \citet{everpresent} and elsewhere derive the effect from purely classical fluctuations. Of course, in order to  deliver a fully quantum theory of gravity, these considerations cannot be but first steps towards the full theory. Nevertheless, the prediction of the everpresent $\Lambda$ on the basis of such simple and straightforward ideas connected to CST surely is intriguing, particularly also in light of the explanatory plight of standard approaches to account for the acceleration of the universe's expansion.

\section{Summing up}
\label{sec:summing}

We could have cast the issue of the emergence of spacetime in CST in terms of Pythagoreanism (as discussed in chapter 1), the view according to which all being is mathematical, formal, or abstract. The worry would be that if our physical world is not fundamentally spatiotemporal, then the Pythagoreans were right and there could not exist something physical, material, or concrete. But this implication does not hold: the world according to CST is quite clearly non-spatiotemporal in its fundamental nature (as we have argued in chapter 2
), and yet, space, time, and perhaps even material objects of our manifest world emerge, or at least \textit{can} emerge, as we hope to have sketched in this chapter. We have articulated what we take the task of establishing this to consist in, and we have invoked spacetime functionalism in order to reject further (and impossible) tasks such as deriving the qualitative nature of spacetime from mathematical structures. With spacetime functionalists, we maintain that a functional reduction of spacetime and its roles in other theories and in their empirical confirmation exhausts the task to be completed. It is by means of this functional reduction, and by means of it alone, that the mathematical derivations involved in the task obtain physical salience.

Apart from this central task of establishing the emergence of spacetime as sketched in \S\ref{sec:necconds} and \S\ref{sec:sptfunctionalism}, we have also addressed foundational and philosophical issues that arise in the context of the emergence of spacetime in CST. We discussed the consequences of CST and its dynamics needed to regain spacetime for the metaphysics of time (\S\ref{sec:becoming}) and the non-locality inherent in CST and the related questions of how to incorporate matter into the framework and thereby identify its empirical signatures (\S\ref{sec:lorentz}). Both of these issues have put the difference and the commonalities between (classical) CST and (continuum) relativistic spacetimes into sharper contrast. But more interestingly still, they show how inextricably connected physics, mathematics, and philosophy are in the pursuit of a theory of QG.

\bibliographystyle{plainnat}
\bibliography{biblio}

\end{document}